\documentclass[12pt,flushrt,preprint]{aastex}
\usepackage[]{natbib}
\usepackage{hyperref}

\newcommand{\id}{ {\rm d} }

\newcommand{\bvec}[1]{ \mbox{\boldmath$#1$} }

\newcommand{\rj}{{ R_{\rm J}}} 
\newcommand{\cs}{{ c_{\rm s}}} 

\newcommand{\ME}{M_{\oplus}}
\newcommand{\RJ}{R_{\rm J}}
\renewcommand{\cs}{c_{\rm s}}

\begin{document}

\shorttitle{}
\shortauthors{}

\title{Forward and Inverse Modeling for Jovian Seismology}

\author{Jason Jackiewicz}
\affil{New Mexico State University, Department of Astronomy, P.O. Box
  30001, MSC 4500, Las Cruces, NM 88003, USA}
\email{jasonj@nmsu.edu}

\author{Nadine Nettelmann}
\affil{Institut f\"{u}r Physik, Universit\"{a}t Rostock, 18051, Rostock, Germany}

\author{Mark Marley}
\affil{NASA Ames Research Center, MS-245-3, Moffett Field, CA 94035, USA}

\author{Jonathan Fortney}
\affil{Department of Astronomy and Astrophysics, University of California, Santa Cruz, CA 95064, USA}

\begin{abstract}

Jupiter is expected to pulsate in a spectrum of acoustic modes and recent re-analysis of a  spectroscopic time series has identified a regular pattern in the spacing of the frequencies \citep{gaulme2011}. This exciting result can provide constraints on  gross Jovian properties and  warrants a more in-depth theoretical study of  the seismic structure of Jupiter. 
With current instrumentation, such as the SYMPA instrument  \citep{schmider2007}  used for the \citet{gaulme2011} analysis,
we assume that, at minimum, a set of global frequencies extending up to angular degree $\ell=25$ could be observed. 
In order to identify which  modes would best constrain models of Jupiter's interior and thus help motivate the next generation of observations, we explore the sensitivity of derived parameters to this  mode set.
Three different models of the Jovian interior are computed and the theoretical pulsation spectrum from these models for $\ell\leq 25$ is obtained. We  compute sensitivity kernels and  perform linear inversions to infer details of the expected discontinuities in the profiles in the Jovian interior.
We find that the amplitude of the sound-speed jump of a few percent  in the inner/outer envelope  boundary seen in two of the applied models should be reasonably inferred  with these particular modes. Near the core boundary where models predict large density discontinuities, the location of such features can be accurately measured, while their amplitudes have more uncertainty.
These results suggest that this mode set would be sufficient to infer the radial location and strength  of expected discontinuities in Jupiter's interior, and  place strong constraints on the core size and mass. We encourage new observations to detect these Jovian oscillations.

\end{abstract}


\maketitle


\section{Introduction}

Characterizing the interior structure of the Solar System's giant planets is critical for understanding
their origin and evolution and the conditions in the protosolar cloud.  Yet key characteristics of the interiors of the giant planets are still unresolved. We do not know the mass of Jupiter's and Saturn's core or
whether the interiors of Uranus and Neptune  are composed of compositionally distinct layers \citep[e.g.,][]{guillot2007}.

Measurements of their gravity fields can be used to discern the internal distribution of mass
within the planets, however, these data are most useful in determining the density profile in only the outer
portions of the planet. For Jupiter and Saturn, uncertainties in the equation of state for hydrogen at high pressures also
substantially complicate efforts to disentangle the density profile within the inner half of the planet.
For these reasons the mass of their present cores is not well constrained. Possible masses range from 0-10  Earth masses for Jupiter \citep{saumon2004,fortney2010,leconte2012}, and 0-20 Earth  masses for Saturn \citep{guillot1999,gudkova1999,saumon2004}. This lack of knowledge of core mass limits our
understanding of the planetary formation mechanism. The NASA Juno mission will in 2016 measure
the gravitational harmonics of Jupiter to high accuracy, but these measurements will be most useful
in constraining the structure of the outer regions of the planet, including for example, the nature of
the deep atmospheric zonal flows. If Juno can measure the tidal response of Jupiter to Io, $k_2$, the
core  mass might be measurable, but this is not assured \citep{hubbard1999}. Even if
$k_2$ is measured, this parameter is highly degenerate with respect to central condensation.  The same uncertainties that allow for different Jupiter models that all match $J_2$ exactly, will also apply to $k_2$. In particular, internal layer boundaries introduce a degeneracy beyond that from uncertainties from the equation of state \citep{nettelmann2011}. Hence, an independent determination of layer boundaries in the deep interior is still required.


With Jovian seismology, we have the unique opportunity to probe our interior modelling approach of
assuming several homogeneous layers with sharp layer boundaries, in contrast to an inhomogeneous,
but smooth density gradient. If this approach should be invalidated for Jupiter, the hottest
giant planet, we might conclude that it fails even more for the colder giant planets (Saturn, Uranus, Neptune) where, in addition, sedimentation is likely to occur.

 It has been recognized for well over thirty years that detection of acoustic oscillations of a giant planet
would provide a powerful probe of its interior structure \citep{vorontsov1976}.  Since that time a number of papers have
investigated the nature of Jupiter's oscillation modes, and a number of attempts 
to detect Jovian oscillations have been made. With the recent detection of the large global-mode spacing by ground-based observations \citep{gaulme2011}, Jovian seismology has just begun to transition from its infancy to becoming a useful tool for constraining interior models. For comparison, helio- and stellar seismology have made tremendous strides in this same time period in both theory and observations, vastly improving our knowledge of the interiors of the Sun and some stars.

Numerous authors, for example, \citet{vorontsov1976,vorontsov1981,marley1991,mosser1991,provost1993,gudkova1999}, have attempted to study the  frequency spectrum of Jupiter and/or Saturn both theoretically and observationally. On the theoretical side, the rapid rotation \citep{vorontsov1981} and structure discontinuities \citep{provost1993} of Jupiter make for a challenging computation,
but those modes which sense the core have frequencies of about 1 to 2~mHz, centered around
1.5~mHz. This frequency range depends fundamentally on the speed of sound in the metallic and
molecular hydrogen envelope of Jupiter. Fortunately, the tools of stellar and solar seismology allow one to understand the best modes for constraining interior models. A better understanding of which particular set of frequencies will provide the greatest amount of information would allow more focused observational searches.

Here we seek to compute an updated set of Jupiter oscillation frequencies for the most recent Jupiter models.  We restrict the theoretical mode set to those with a spherical harmonic angular degree $\ell$ up to 25 and frequencies below the acoustic-cutoff frequency. The choice for the maximum degree is roughly  based on  the spatial  resolution of current instrumentation. Given this mode set, we then seek to establish what physical conditions in Jupiter's interior could be reasonably inferred using forward and inverse seismic techniques and given a few simplifying assumptions. 

The paper is organized as follows: The recently computed models are described in \S\ref{sec:models}, and their seismic properties are discussed in \S\ref{sec:seis}. \S\ref{sec:forward} shows the computation of the forward problem, and a full description of the inverse problem,  the solution that we choose to employ, and results of example inversions using the given modes is presented in \S\ref{sec:inverse}. Finally, in \S\ref{sec:conc}, we summarize our main conclusions.

\section{Jovian Models}
\label{sec:models}

We have calculated the interior profiles of three different Jupiter models. The models are chosen to encompass the range of current knowledge.  We assume the existence of a core, and explore core sizes up to the maximum currently allowed. Within the envelope, density discontinuities may occur, for instance, due to helium sedimentation. A discontinuity in the abundance of heavy elements may also occur at this location.
The models are based on physical equations of state (EOS) for hydrogen and helium from different sources.
In the following, we describe those models in detail.


\subsection{Model descriptions}

\subsubsection*{The SCV and LMR models}

Models LMR and SCV  are computed by the procedure described in \citet{nettelmann2008,nettelmann2012}. They make use of the standard three-layer structure type, 
meaning two convective, homogeneous, and adiabatic envelopes above a core of heavy elements, 
where the abundances of helium and heavy elements in the hydrogen rich envelopes are 
chosen to reproduce observational constraints as follows. 

The He abundance in the outer envelope, which begins at the 1~bar level with temperature $T_1$,
is fixed to the Galileo entry probe value of $Y_1=0.238 (1-Z_1)$, 
where $Z_1$ is the heavy element mass fraction there. In the inner envelope, 
the helium abundance, $Y_2$, is adjusted to give an overall H-He mass fraction of 0.275 in agreement with the value of the protosolar cloud out of which the giant planets in the solar system  formed. The abundances of elements that did not condense nor chemically bind to dust under protosolar cloud conditions should still be conserved in Jupiter, on average.  The respective heavy element abundances in the envelopes, 
$Z_1$ and $Z_2$, are chosen to give an internal density profile that is consistent with the 
measured gravitational moments $J_2$ and $J_4$ within the observational error bars \citep{guillot2005}.

While the position of the transition between the envelopes, parametrized by a transition
pressure $P_{1-2}$ influences the resulting solutions, especially the core mass $M_{c}$, 
it is actually not fixed by an observable.
In particular, we have $P_{1-2}=2$~Mbar, $M_{c}=1.28\:\ME$, $J_4/10^{-4}=-5.84$, $T_1=165$~K, 
$Y_2=0.2593$, $Z_1=0.0793$ and $Z_2=0.0817$ for SCV, and $P_{1-2}=8$~Mbar, $M_{c}=3.57\:\ME$, $J_4/10^{-4}=-5.89$, $T_1=170$~K, $Y_2=0.3112$, $Z_1=0.0381$, and $Z_2=0.1283$ for  LMR. The LMR model is identical to model J11-8a of \citet{nettelmann2012}.

Most importantly, models SCV and LMR differ in the equations of state (EOS) used for H, He, and metals in the envelopes. For model SCV we used the SCvH EOS with H-SCvH-i for hydrogen, He-SCvH for helium \citep{saumon1995}, and this He EOS scaled in density by a factor of four to represent 
heavy elements. For model LMR we used LM-REOS \citep{nettelmann2008,nettelmann2012} with water representing
metals. The pressure-density relation in the cores was obtained from the rock EOS by \citet{hubbard1989}.

To get smooth derivatives d$P$/d$\rho$ with model LMR, the planetary  $P(\rho)$ profile 
in each layer was reduced to a small number ($\sim 10-40$) of points $P_i$, transformed to an equidistant grid in  log$P$ on which then the derivatives (dlog$\rho$/dlog$P$)$_{P_i}$ were taken by the 4-point Richardson method,
and related to the radial coordinate values by spline or linear interpolation so that finally,
$c_s=\sqrt{(P/\rho)\,{\rm dlog}\rho/{\rm dlog} P}$,  could be calculated.

\subsubsection*{The MH08 model}
Model MH08  is the Jupiter model of \citet{militzer2008}, with a 16$\ME$ core of ices and rocks, $Y_1=0.238$ and $Z_1=0.014$ throughout the single envelope, and $J_4/10^{-4}=-6.14$. 
The numerical derivatives along the published tabulated profile turned out sufficiently smooth.

\begin{figure}
  \centering
  \centerline{\includegraphics[width=\textwidth]{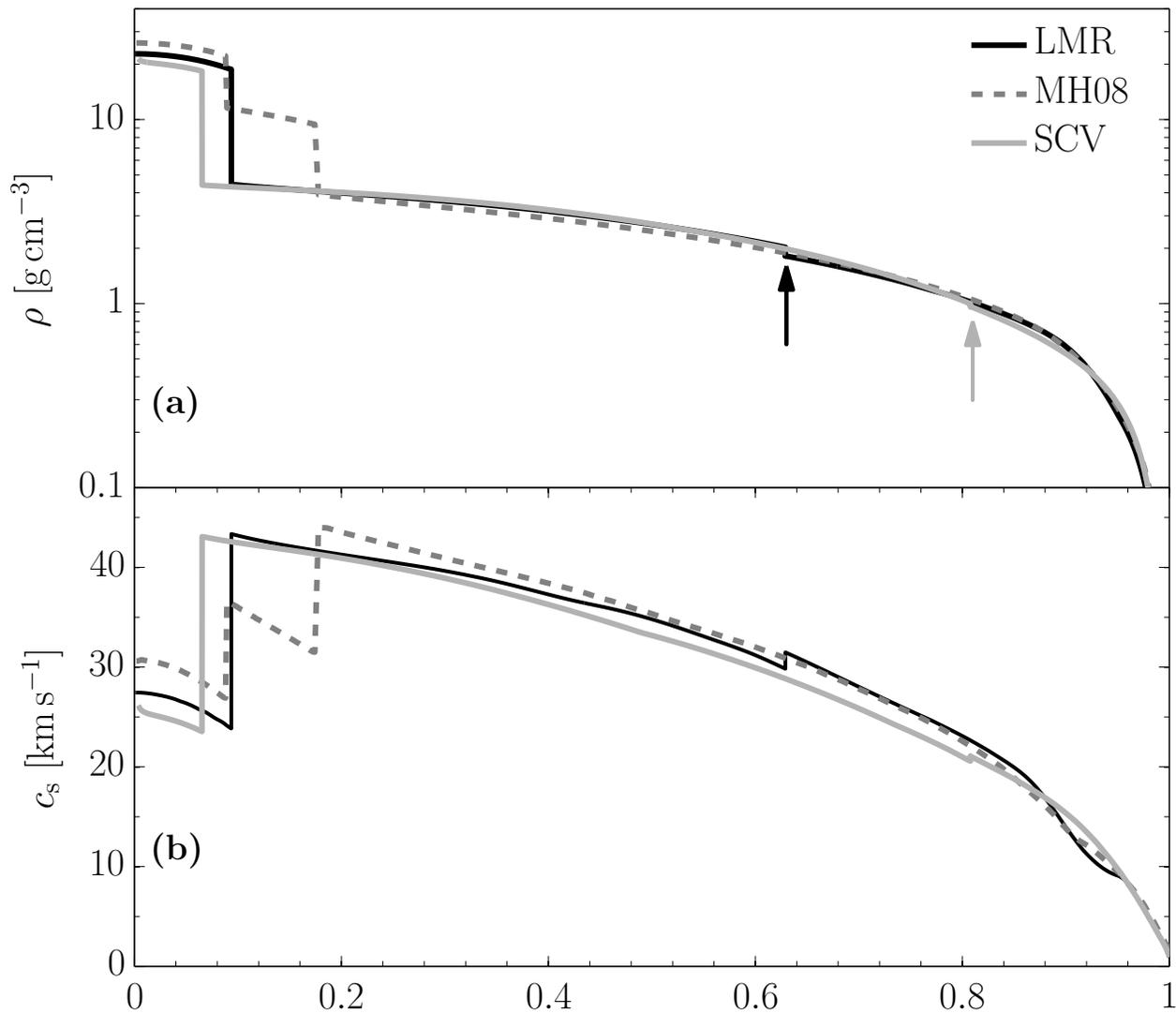}}
  \vspace{-.6\textwidth} 
  \centerline{\large\bf\hspace{0.12\textwidth}(a)\hfill}
  \vspace{.35\textwidth}
  \centerline{\large\bf\hspace{0.12\textwidth}(b)\hfill}
  \vspace{.1\textwidth}
  \caption{(a) Density and (b) sound-speed profiles for the three models as a function of scaled Jupiter radius, $R_{\rm J}$. Noted by  arrows are small jumps in density for the LMR model at $\sim 0.63\RJ$ and for the SCV model at $\sim 0.8\RJ$. Similar features are also present in the sound speed.}
  \label{fig:cs}
\end{figure}


\subsubsection*{Summary of models}
To our best knowledge, these three models represent the widest range of currently offered 
Jupiter models with physical EOS.
They span the observational uncertainty in the 1~bar temperature (165$-$170~K);
differ in core composition (pure rocks or ice-rock); include models with a small core 
(LMR) and a large core (MH08); have a strong, small, or absent layer boundary in the
envelope;  assume a helium enriched deep envelope (SCV, LMR) or not (MH08); assume 
that differential rotation has a negligible effect on $J_4$ (SCV, LMR) or a significant 
effect (MH08), have atmospheric metallicities in agreement with the measured abundances
(MH08) or with the prediction that oxygen is $\sim 2-4$ times solar (SCV, LMR); and, finally,
include H-He mixing effects in the EOS (MH08) or not (SCV, LMR).

\subsection{Density and sound-speed profiles}

Figure~\ref{fig:cs} shows radial profiles of the sound speed and density of the three models. In homogeneous layers, the sound speed $\cs(r)$ rises steadily with depth as the interior becomes warmer and increasingly degenerate. When dissociation occurs,  the heat capacity at constant volume, $C_V$ rises more than $C_P$, the heat capacity at constant pressure, and the rise of $\cs(r)$ with depth is delayed. This effect can be seen at around 0.9\,$\RJ$ in the MH08 and LMR models. The inner/outer envelope transitions of the LMR and SCV models are denoted by arrows in Fig.~\ref{fig:cs}, and will be the targets of our inversion analysis in the following sections. At the deepest layer boundaries, the steep rise in mass density maps onto a steep decrease in $\cs$. 
The lower sound velocities in the deep interior of Jupiter as predicted by models SCV and LMR
compared to MH08 are a consequence of higher envelope metallicities, hence higher mass 
densities there.

\section{Model seismic parameters}
\label{sec:seis}

Model profiles permit the computation of the expected oscillations in Jupiter, which are likely  acoustic in nature and driven by convection, similar to the case of the Sun and solar-like stars. The oscillations are expected to couple to the troposphere and produce small-amplitude albedo variations \citep{mosser1995,gaulme2005} that in principle could be detected with appropriate observations.  In what follows, we ignore rotational effects on the frequencies of the pulsation modes, whose effects have been studied previously \citep[e.g.,][]{vorontsov1981,mosser1990,lee1993}. While rotation imparts non-negligible shifts in the frequency spectrum for non-radial pulsations (on the order of $27\,{\rm\mu Hz}$), our goal here is to provide an overall study on the usefulness of the pulsations on structure determination and these small deviations will not affect our main conclusions.

\subsection{Adiabatic oscillations}
\label{sec:modes}

We consider a linear theory of adiabatic, radial and non-radial oscillations. Computations are carried out using small perturbations to the ordinary differential equations of stellar structure, following the general formulation found in \citet{unno1989}. For each model, we obtain the eigenfunctions and eigenfrequencies $\omega^0_{n\ell}$ for any given angular degree $\ell$ and radial order $n$. The superscript ``0'' distinguishes model frequencies from observed ones, as detailed later.  In this study, we truncate the algorithm at $\ell\leq 25$. This angular degree limit is consistent with what could be obtained, in principle, with the recently built and employed SYMPA spectrograph \citep{schmider2007} used for several ground-based observing campaigns. Of course, higher-degree modes should certainly be possible to detect from space on appropriate instruments with sufficient spatial resolution of the Jovian disk. For comparison, the Sun reveals degrees up to several thousand in high-resolution images.

\begin{figure}
  \centering
  \includegraphics[width=\textwidth]{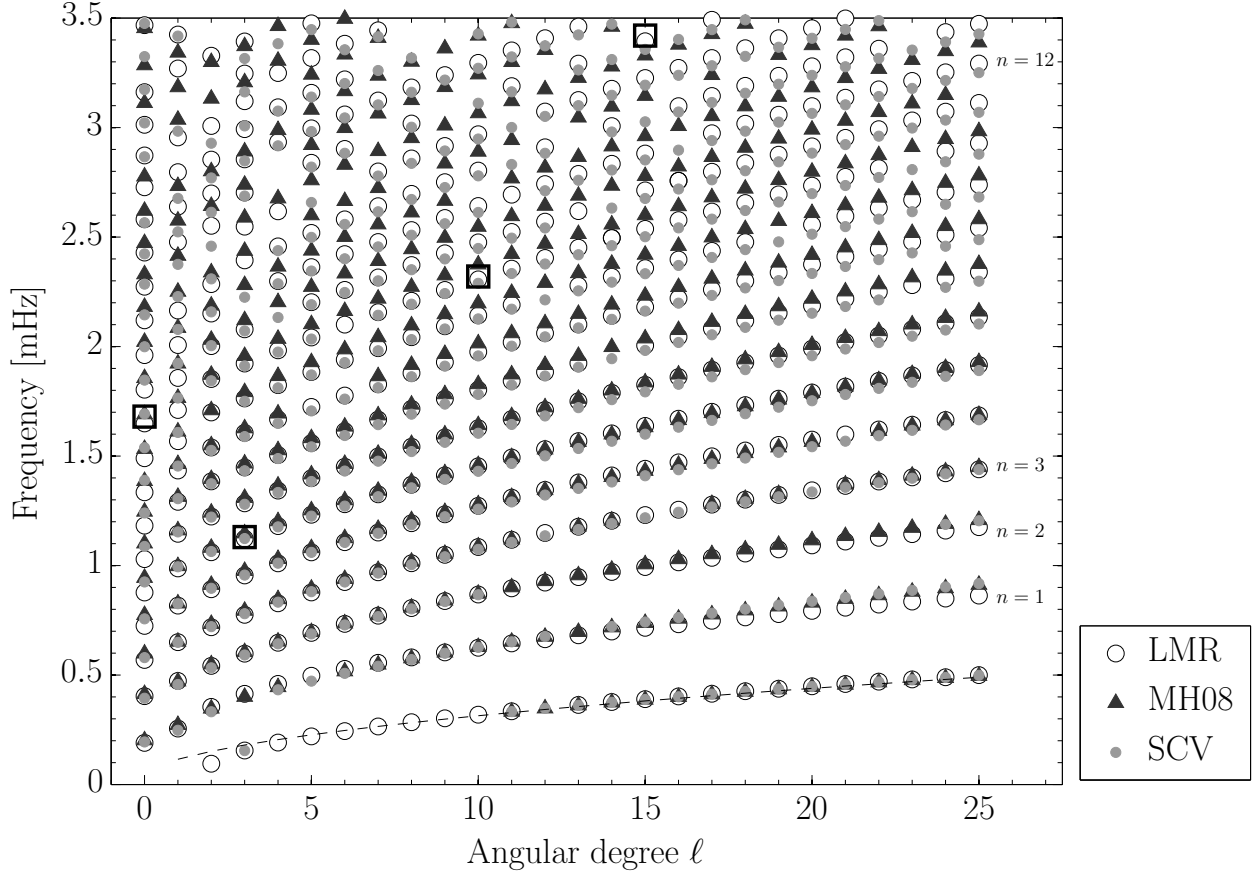}
  \caption{Computed frequencies $\nu_{n\ell}^0$  for each of the three models in $\ell - \nu$ space for $\ell\leq 25$ and $\nu\leq 3.5$~mHz. Several radial orders ($n$) are labeled for reference. The dashed line along the lowest ridge denotes the dispersion relation for the surface-gravity (fundamental $n=0$) mode. Black squares highlight the four modes for which kernels are plotted in Fig.~\ref{fig:kerns}.}
  \label{fig:lnu}
\end{figure}

Figure~\ref{fig:lnu} shows the theoretically computed cyclic frequency $\nu^0_{n\ell}=\omega^0_{n\ell}/2\pi$ as a function of the angular degree $\ell$ for each of the three models. The lowest frequency ``ridge'' is  the surface-gravity (fundamental) mode, which is not an acoustic mode and has a dispersion relation $\omega = \sqrt{g k}$, where $\omega$ is the angular frequency, $g$ is surface gravitational acceleration, and $k$ is wavenumber: $k^2\RJ^2 = \ell(\ell+1)$. The fundamental mode is analogous to deep ocean waves \citep{lighthill1978}.  The rest of the spectrum shows the acoustic pressure modes up to $\ell=25$. The maximum frequency we consider is $3.5$~mHz, which is chosen as a value slightly higher than the expected  atmospheric acoustic cutoff frequency  for Jupiter \citep{mosser1995}.  Physically, modes with higher angular degree are more sensitive to the outer envelope of the planet, while those with low degree probe the core. Increasing frequencies for a given angular degree denote successive radial orders $n$. The highest frequency modes correspond to roughly  $n\approx 22$.   The $\ell=0$ modes are the radial solutions, whose eigenfunctions have no nodes on the surface.

It is unknown whether all modes that we expect to exist in Jupiter based on interior models will in fact be excited, or excited strongly enough to reach detectable levels. Convective mode-excitation mechanisms are complex and a full understanding of all the processes responsible, including damping, is still lacking.

We finally note that the solution finder of the differential equations is sensitive to the input frequency guess, and  a solution is not always found for a given $\ell$ since the input grid needs to be sufficiently fine. Thus, for  different models   there are occasionally  ``missed'' modes, as one notices in  Fig.~\ref{fig:lnu}.  As the equations are adiabatic, we also cannot provide the expected amplitudes of any pulsation. We point out, however, that the recent observational work by \citet{gaulme2011} found  that the highest amplitude mode had a frequency of about $\nu=1.2$~mHz. This seems  plausible, as recent work for the Sun and solar-like stars has shown a robust relationship between the frequency at maximum power (so-called $\nu_{\rm max}$) and the acoustic cutoff frequency \citep{belkacem2011}. In the case of Jupiter, $\nu_{\rm max}$ should  therefore be about one half of the expected acoustic cutoff frequency of $\nu\approx 3$~mHz \citep{mosser1995}.


\subsection{Large frequency separation}

Various diagnostics computed from the seismic frequency spectrum are very useful to understand global properties that could be obtained from observations. To leading order for low-degree modes of high radial order, acoustic oscillations in convectively unstable bodies can be shown to follow a regular pattern given by
\begin{equation}
  \nu_{n\ell}\simeq \left(n+\frac{\ell}{2} + \frac{1}{4}+\alpha\right)\Delta\nu,
  \label{nunl}
\end{equation}
where  $\alpha$ is a function capturing effects of near-surface regions \citep[see][and references therein]{aerts2010}. The quantity $\Delta\nu$ is the  large frequency separation, and is the inverse of twice the sound travel time from the planet center to the surface:
\begin{equation}
  \Delta\nu = \left[ 2\int_0^\rj\frac{\id r}{\cs} \right]^{-1}.
  \label{eq:lsep}
\end{equation}
The large frequency spacing quantifies the expected asymptotic spacing (for large radial orders $n$, i.e., high frequencies) between pairs of modes $(n,\ell)$ and $(n+1,\ell)$, and scales with the square root of the mean density of the planet. The regular pattern  in the asymptotic limit is somewhat visible by eye in the frequency spectrum in Fig.~\ref{fig:lnu}.

\begin{figure}
  \centering
  \centerline{
  \includegraphics[width=.33\textwidth]{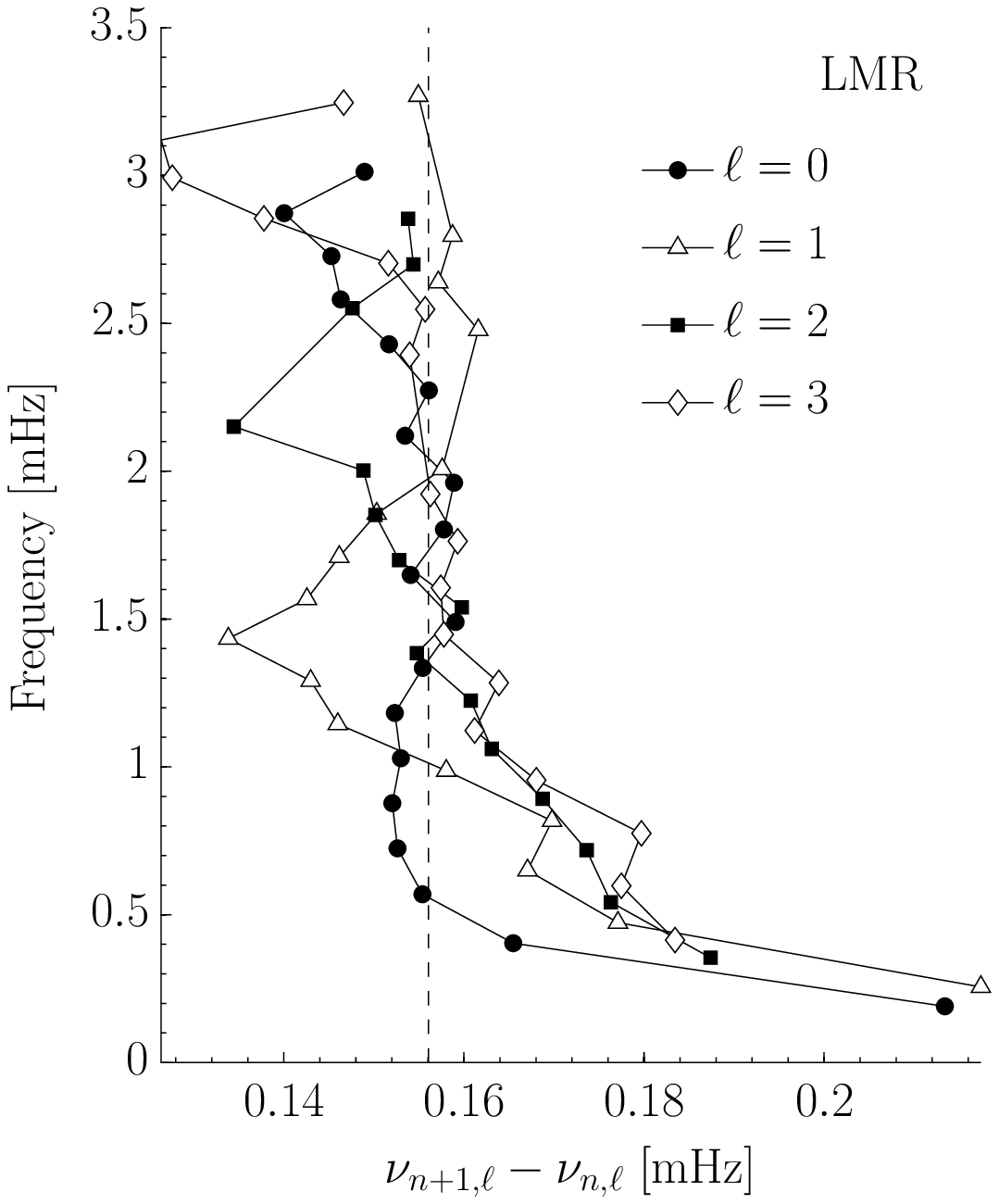}
  \includegraphics[width=.33\textwidth]{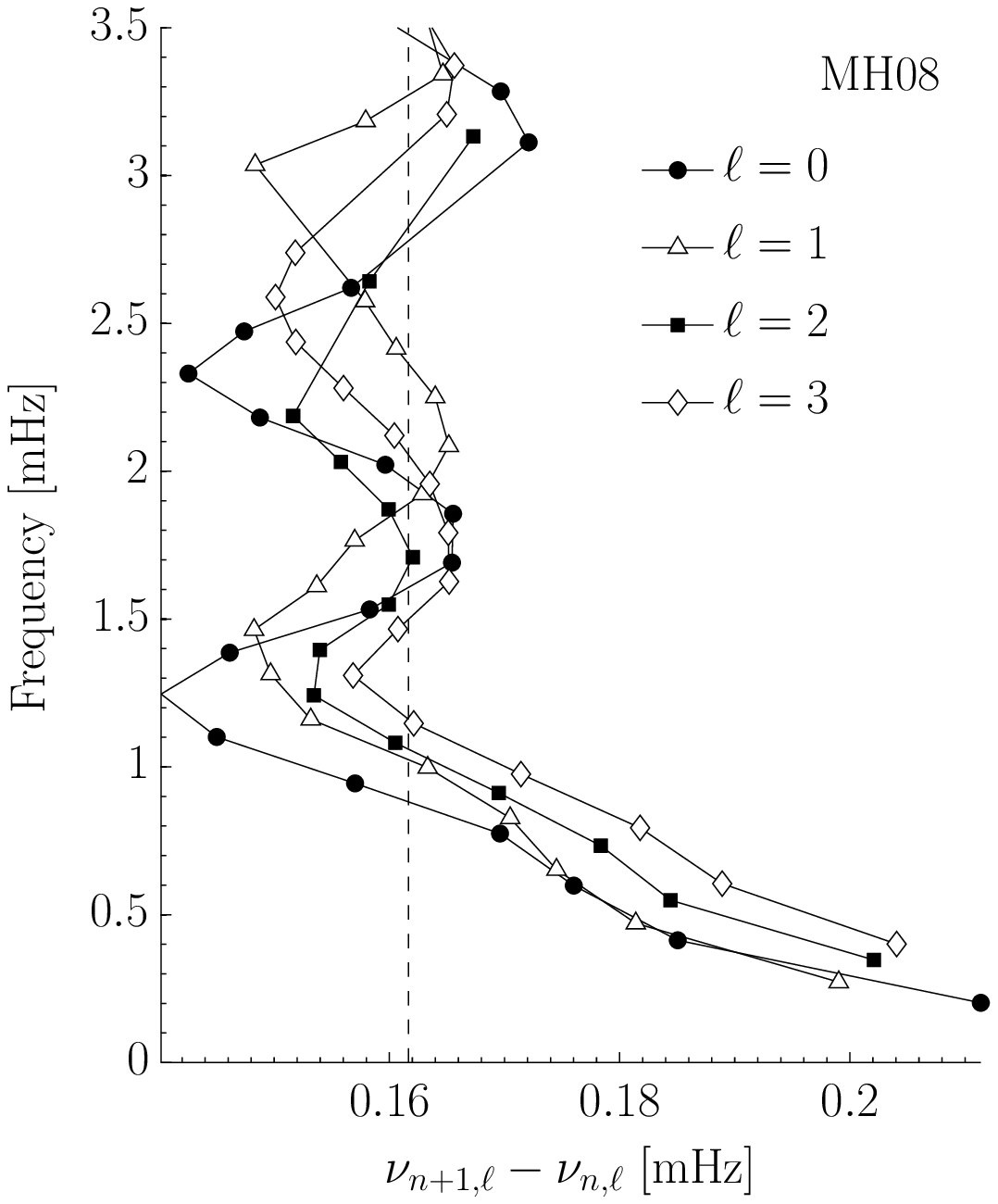}
  \includegraphics[width=.33\textwidth]{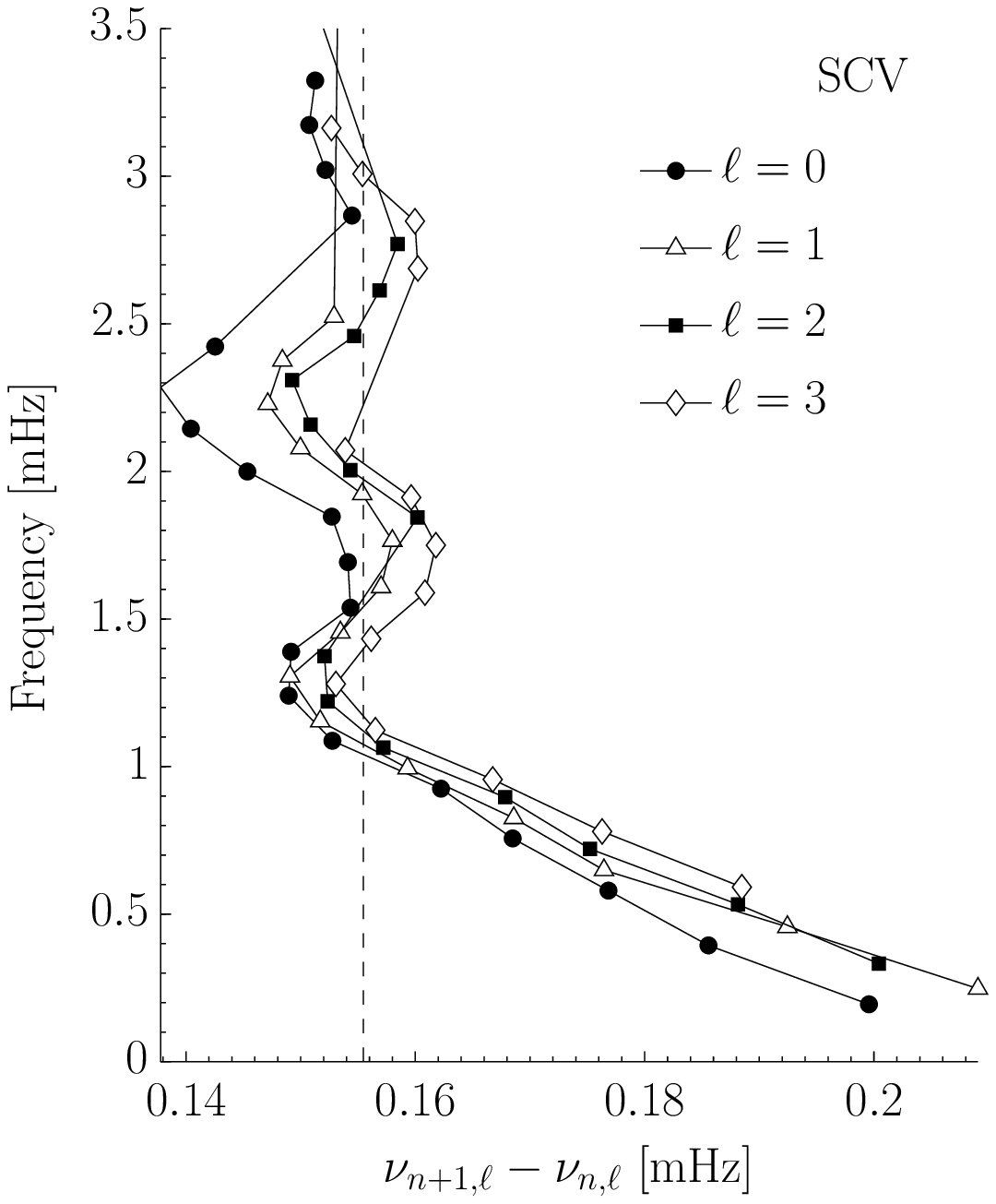}}
  \caption{Large frequency separation $\Delta\nu$ of low-degree modes for each of the three models. The vertical dashed line is the theoretical value of $\Delta\nu$ for a given model from  Eq.~\ref{eq:lsep}.}
  \label{fig:lsep}
\end{figure}

The large frequency separation for the computed modes, along with the expected theoretical value from Eq.~\ref{eq:lsep} are shown in Fig.~\ref{fig:lsep}. For model LMR, we find $\Delta\nu=156.1\,{\rm \mu Hz}$, for MH08,  $\Delta\nu=161.7\,{\rm \mu Hz}$, and for SCV,  $\Delta\nu=155.5\,{\rm \mu Hz}$.  Asymptotic scaling relations for the Sun and solar-like stars point to  a  large frequency separation $\Delta\nu=\Delta\nu_\odot\, (M/M_\odot)^{0.5}\,(R/R_\odot)^{-1.5}$ \citep{brown1991}. Given that Jupiter is about 1000 times less massive and about 10 times smaller in radius, one might expect the  large frequency separation for Jupiter to be roughly similar to that of the Sun: $\Delta\nu_\odot \sim 135\,{\rm\mu Hz}$. The observations by \citet{gaulme2011} found $\Delta\nu\approx 155.3\pm 2.1\,{\rm\mu Hz}$ for Jupiter, indeed very close to the three models studied here.


\begin{figure}
  \centering
  \centerline{
  \includegraphics[width=.33\textwidth]{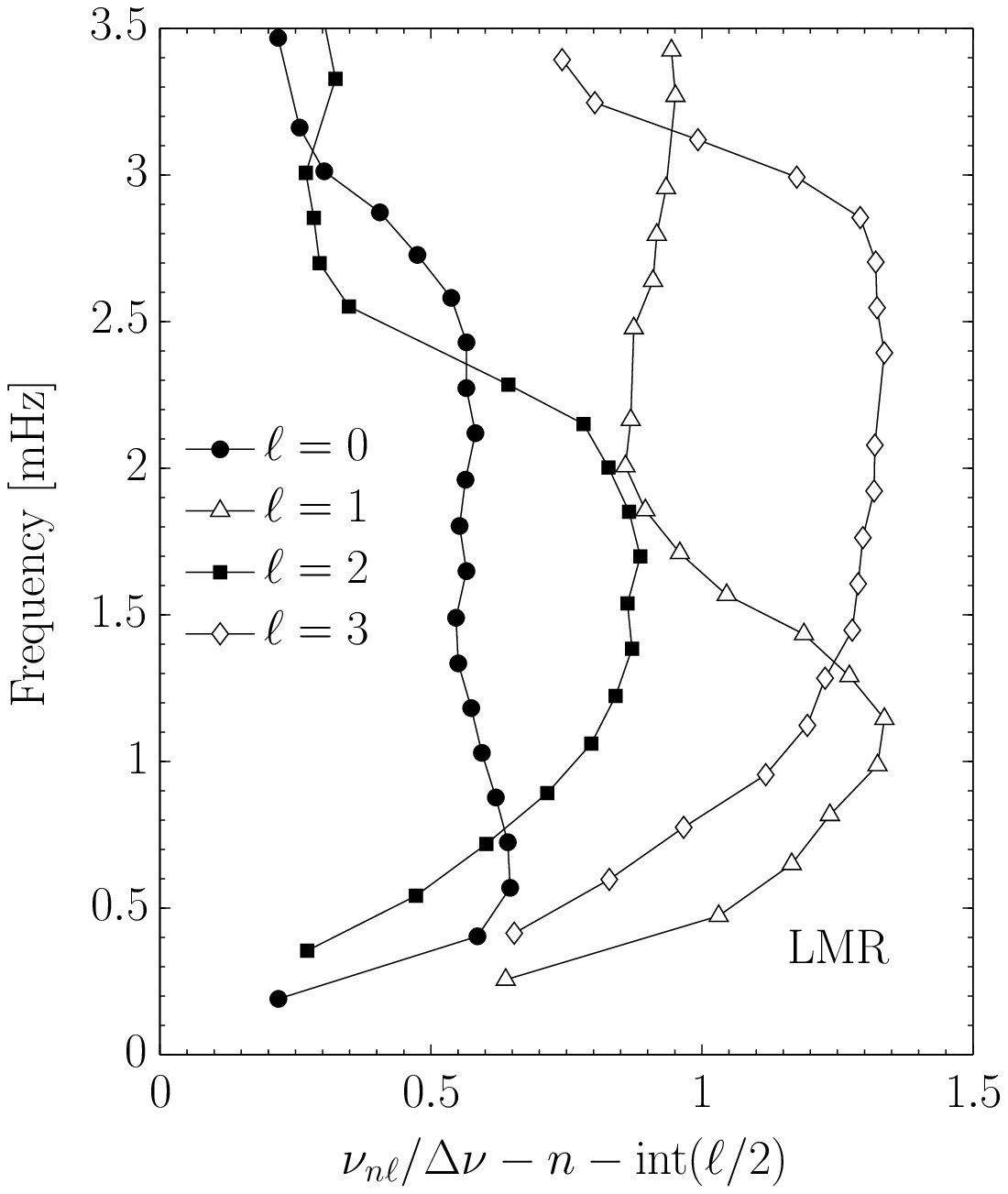}
  \includegraphics[width=.33\textwidth]{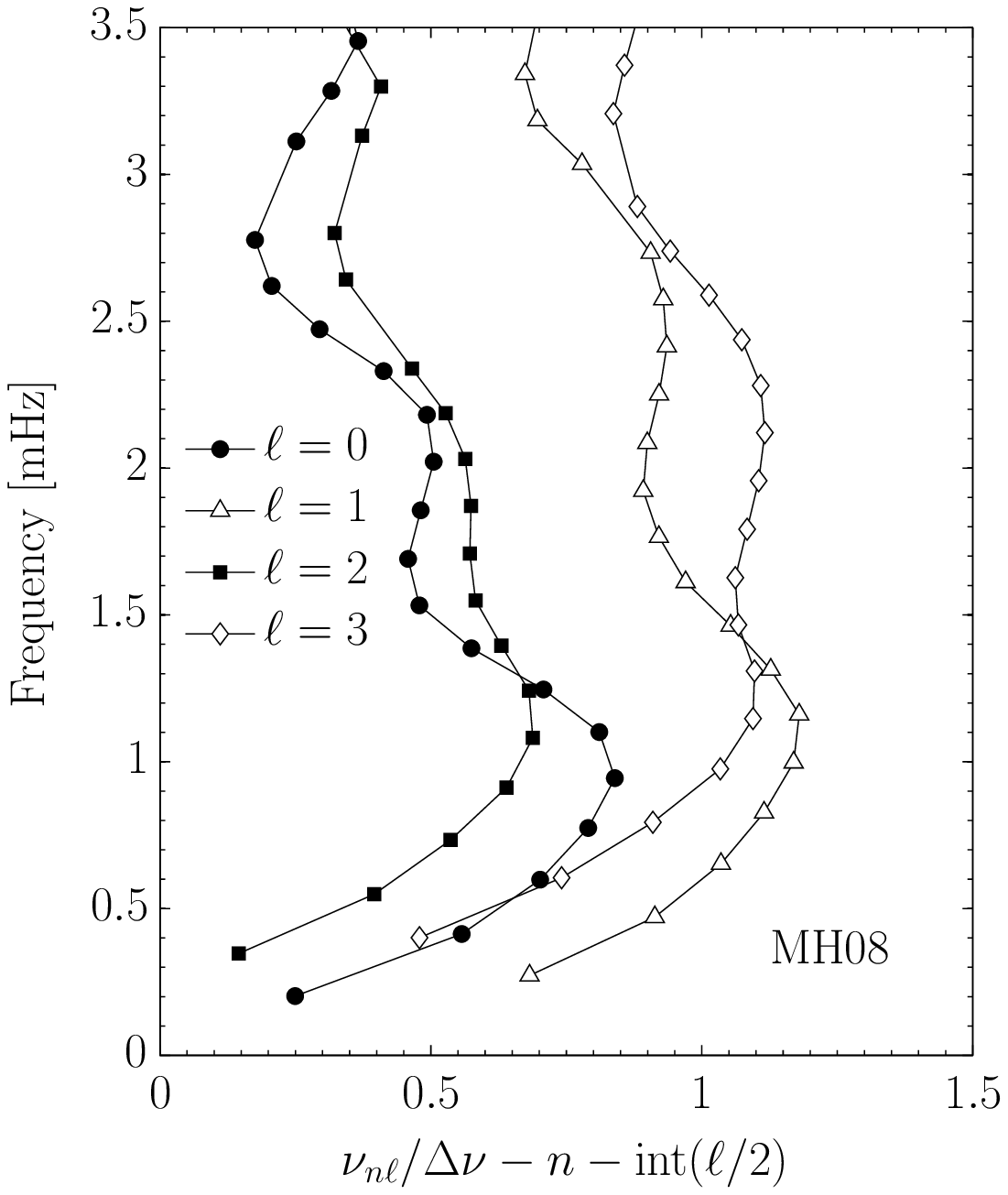}
  \includegraphics[width=.33\textwidth]{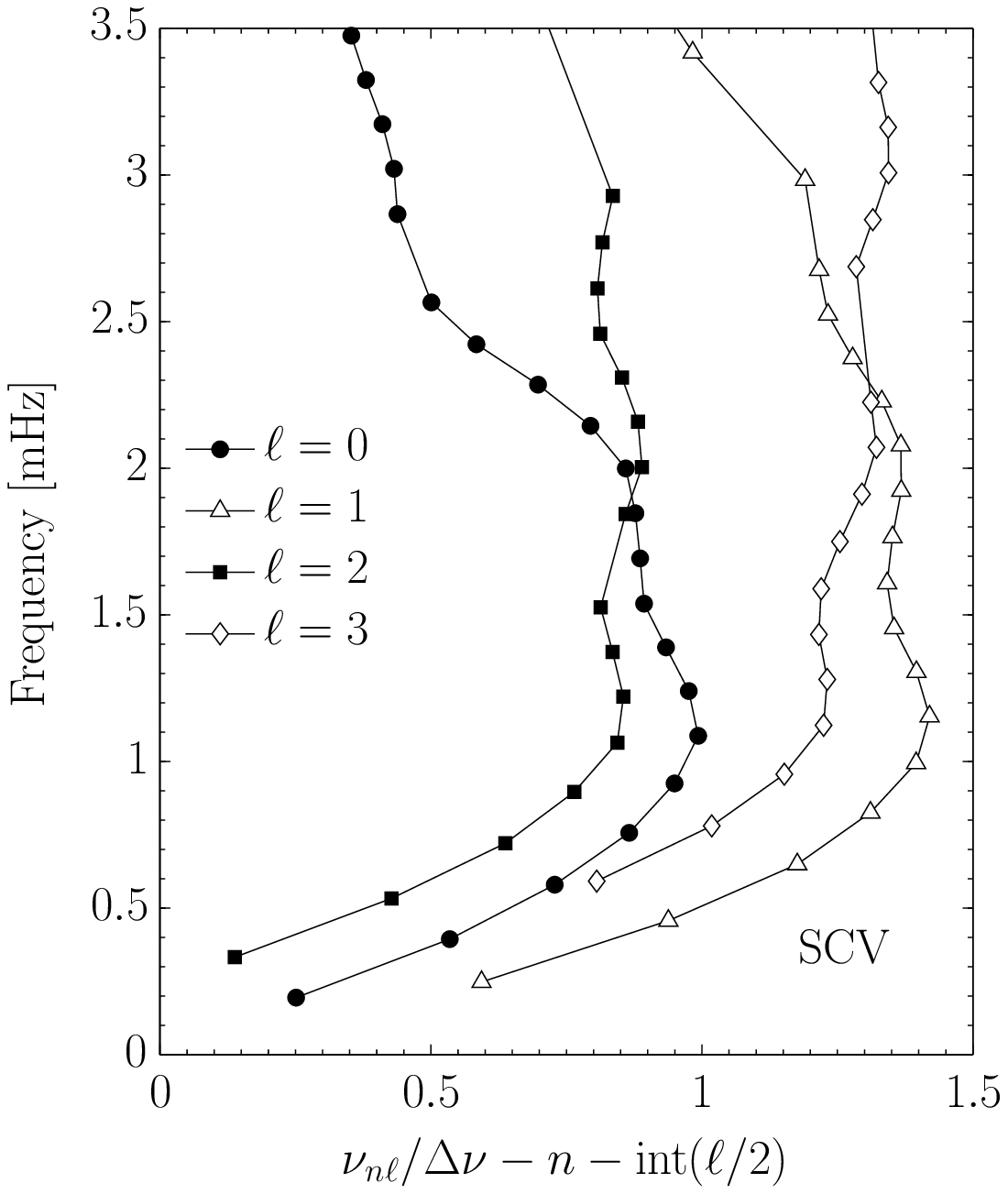}}
  \caption{\'Echelle diagrams from the computed linear adiabatic modes for each of the three models. The symbols represent the different (low) angular degree of the modes. These plots can be thought of as stacked sections of roughly length $\Delta\nu$ of the frequency spectrum.}
  \label{fig:ech}
\end{figure}

Due to the equal spacing between modes of consecutive radial orders of the same degree predicted by asymptotic theory, vertical alignment of modes is expected if plotted in an \'echelle diagram, as is commonly done using frequency observations of stars.  Figure~\ref{fig:ech} shows such diagrams, where the frequency spectrum is vertically stacked from sections of a given length in frequency determined primarily by $\Delta\nu$. As can be seen, Jovian models do not generally demonstrate this vertical behavior for high frequencies  \citep[as first noted by][]{vorontsov1989}, likely due to very strong core contributions to the eigenfunctions from the low-degree  modes and sharp density jumps. The overall features of the \'echelle diagrams, particularly the wiggles and the flattening at low frequencies, are similar to what has been seen in previous Jovian models \citep{provost1993,gudkova1999}.

\subsection{Small frequency separation}

\begin{figure}
  \centering
  \centerline{
  \includegraphics[width=.33\textwidth]{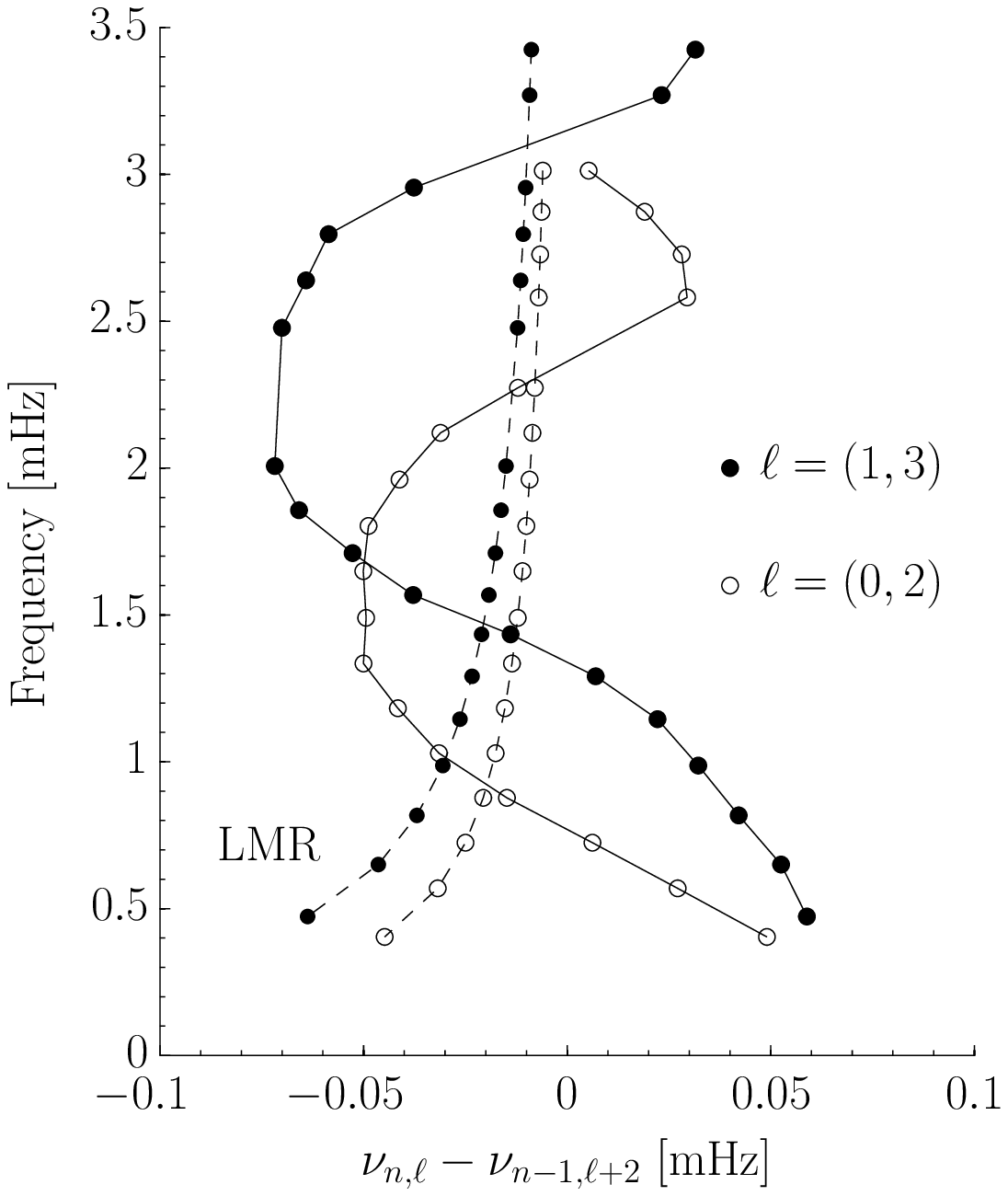}
  \includegraphics[width=.33\textwidth]{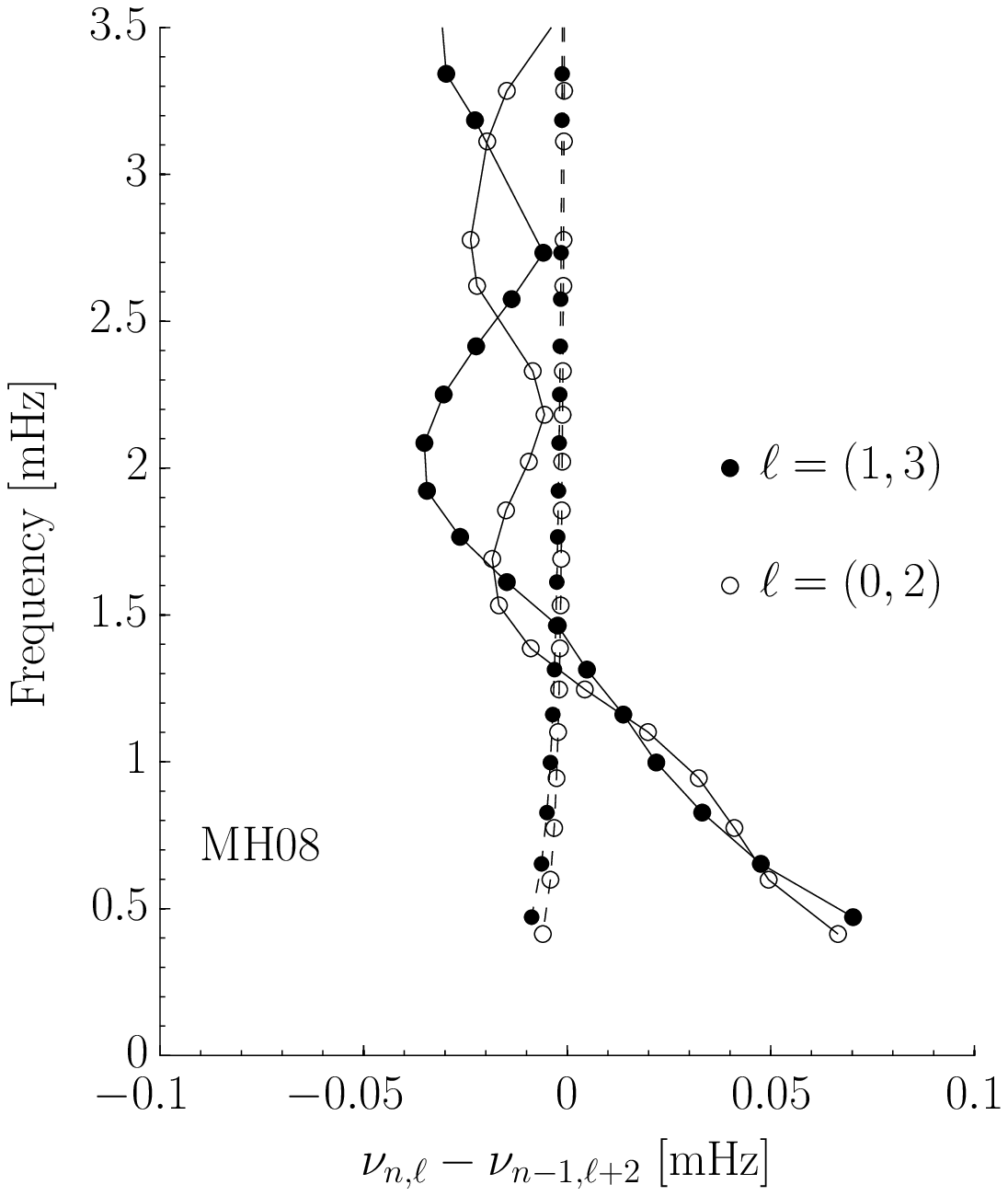}
  \includegraphics[width=.33\textwidth]{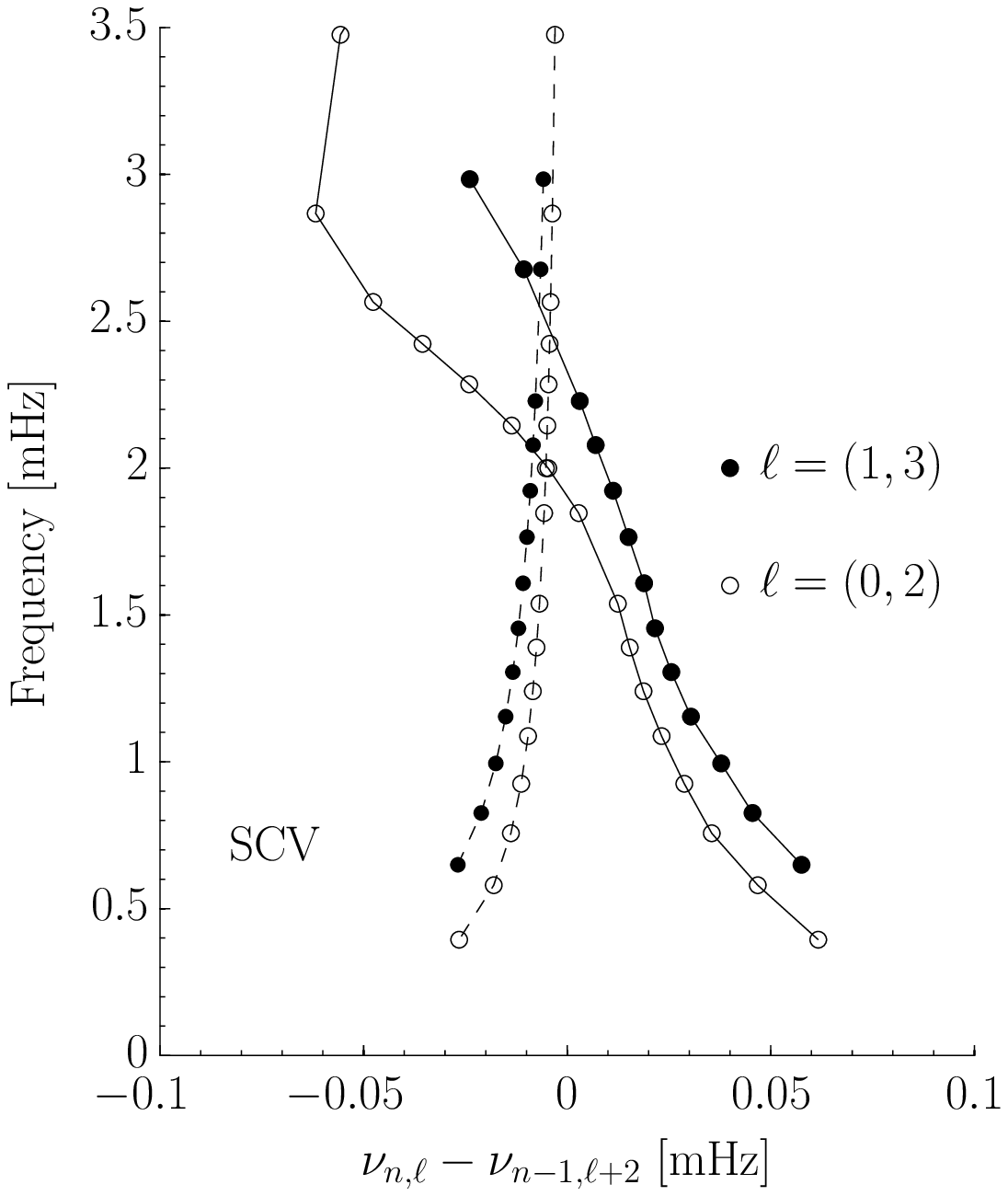}}
  \caption{Small frequency separation $\delta\nu$ for each of the three models. Dashed lines are theoretical values (from Eq.~\ref{eq:smsep}) and solid lines are computed from the model frequency spectrum. The angular degree pairs for consecutive orders are (1,3) and (0,2). Note the weak frequency dependence of the theoretical values from the term in the denominator in  Eq.~\ref{eq:smsep}.}
  \label{fig:smsep}
\end{figure}

From Eq.~\ref{nunl} it is apparent that frequencies of pairs of modes of  $(n,\ell)$ and $(n-1,\ell+2)$ should be similar in value. Expanding to second order in the asymptotic low-degree regime provides a correction to the mode frequency $\nu_{n\ell}$  \citep{tassoul1980}. This results in a breaking of the degeneracy of these pairs, and gives what is known as the small frequency separation
\begin{equation}
  \delta\nu\equiv\nu_{n\ell}-\nu_{n-1\,\ell+2}\approx - (4\ell+6)\frac{\Delta\nu}{4\pi^2\nu_{n\ell}}\int_0^\rj\frac{\id\cs}{\id r}\frac{\id r}{r}.
  \label{eq:smsep}
\end{equation}
 This combination isolates contributions from the sound speed gradient in the integrand, which  is most pronounced at the core-mantle boundary for the case of Jupiter, and therefore provides a strong diagnostic for that region. The pairs of modes we consider  are the lowest degrees $\ell=(0,2)$ and $\ell=(1,3)$ for consecutive orders. The computed separations from the model frequency spectra and the expected separations from Eq.~\ref{eq:smsep} are compared in Fig.~\ref{fig:smsep}. The model separation is seen to span a range of about $100\,{\rm \mu Hz}$ and it is again evident that Jupiter does not simply follow the expected behavior predicted by asymptotic theory. The small frequency separation is also on the order of rotational effects on non-radial modes.

\subsection{Probe depths of the acoustic modes in Jupiter}

\begin{figure}
  \centering
  \includegraphics[width=\textwidth]{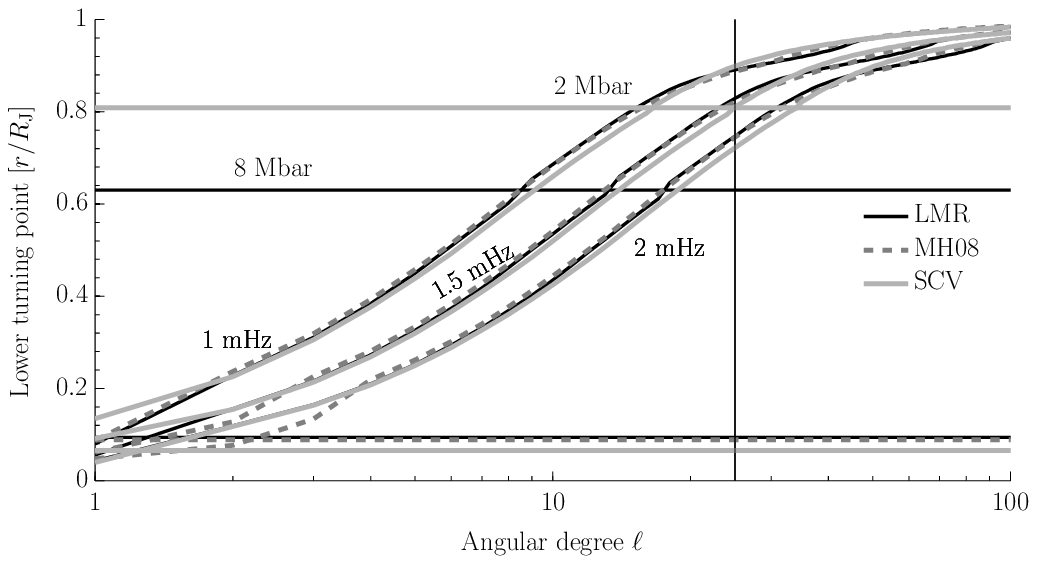}
  \caption{The  lower turning points of a set of modes as a function of angular degree for each of the three models and for three different representative frequencies computed from Eq.~\ref{ltp}. The horizontal lines deep inside Jupiter denote the boundary of the core/metallic region for each model, while the horizontal lines  in the outer $\sim 40\%$ represent the approximate inner/outer envelope boundary. Associated pressures are also labeled for the outer envelope boundaries (model MH08 does not have one). The  vertical  line marks the boundary of the modes considered in this study  ($\ell\le 25$).}
  \label{fig:rt}
\end{figure}

Modes excited in the convective envelope probe into the planetary interior to depths based on frequency and wavenumber, and refract back towards the surface due to the increasing density. These trajectories can be thought of as simple ray paths. To guide the inversion input, we show  in Fig.~\ref{fig:rt} the lower turning point $r_{\rm t}$ for modes of varying angular degree and frequency for all three models. The lower turning point can be understood as the region above which a particular mode is trapped beneath the surface. It is found through the relation 
\begin{equation}
  \frac{c_{\rm s}^2(r_{\rm t})}{r_{\rm t}^2} = \frac{\omega^2}{\ell(\ell+1)},
  \label{ltp}
\end{equation}
where the angular frequency $\omega =2\pi\nu$.  Equation~\ref{ltp} shows that, for a given $\ell$, higher-frequency modes probe deeper than low frequency ones, as  Fig.~\ref{fig:rt} confirms.  High-degree modes are trapped in the outer layers. In the analysis that follows, we exploit the basic property of seismology: that  different modes sense different depths in the interior of the planet, to do targeted studies of distinct interior regions.


\section{Forward modeling - Sensitivity kernels for Jupiter}
\label{sec:forward}

Jovian oscillation frequencies are related to  internal properties such as sound speed and density structure  in a rather complicated and non-linear fashion. As is done in the cases of terrestrial and solar seismology to approximate such a relationship, we perform a linearization about an initial reference model, or three models, as is the situation here \citep[for full details in the solar context, see][]{jcd2002}. For example, let $\rho_0$ and $c_0$ denote the background density and sound-speed profiles of one of the models. Such a model permits calculation of a spectrum of adiabatic oscillations frequencies $\omega_{n\ell}^0$, as already depicted in Fig.~\ref{fig:lnu}. Now let us assume for the moment that several observations of Jovian frequencies are available, $\omega_{n\ell}^{\rm obs}$. We are interested in seeking corrections to the density, $\delta\rho = \rho-\rho_0$, and squared sound speed, $\delta c^2=c^2-c_0^2$, that match the differences between the observed and modeled frequencies, $\delta\omega_{n\ell}=\omega_{n\ell}^{\rm obs} - \omega_{n\ell}^0$. Assuming small differences to those quantities, a linear relation such as
\begin{equation}
   \frac{\delta\omega_{n\ell}}{\omega_{n\ell}} = \int_0^{\RJ}\left[ K_{n\ell}^{\cs^2,\rho}(r)\frac{\delta c^2_s}{c^2_s}(r) + K_{n\ell}^{\rho,\cs^2}(r)\frac{\delta\rho}{\rho}(r)\right]\,\id r + \epsilon_{n\ell}, 
\label{csrho} 
\end{equation} 
can be found.  The quantities  $K$ are known as  sensitivity functions, or kernels, and are computed  using the model profiles and mode eigenfunctions. They are derived by straightforward first-order perturbation theory applied to the basic equations of hydrodynamics. Details and explicit expressions can be found in \citet{gough1991}. The sensitivity kernels relate the interior conditions to the frequency shifts $\delta\omega$. The first kernel above is for squared sound-speed differences ($\delta \cs^2$) without density differences, and the second is for density differences ($\delta\rho$) without sound-speed differences \citep[e.g.,][]{jcd2003}. Relative uncertainties in each observed frequency due to realization noise  are captured by $\epsilon_{n\ell}$.

\begin{figure}
  \centering
  \centerline{
    \includegraphics[width=.5\textwidth]{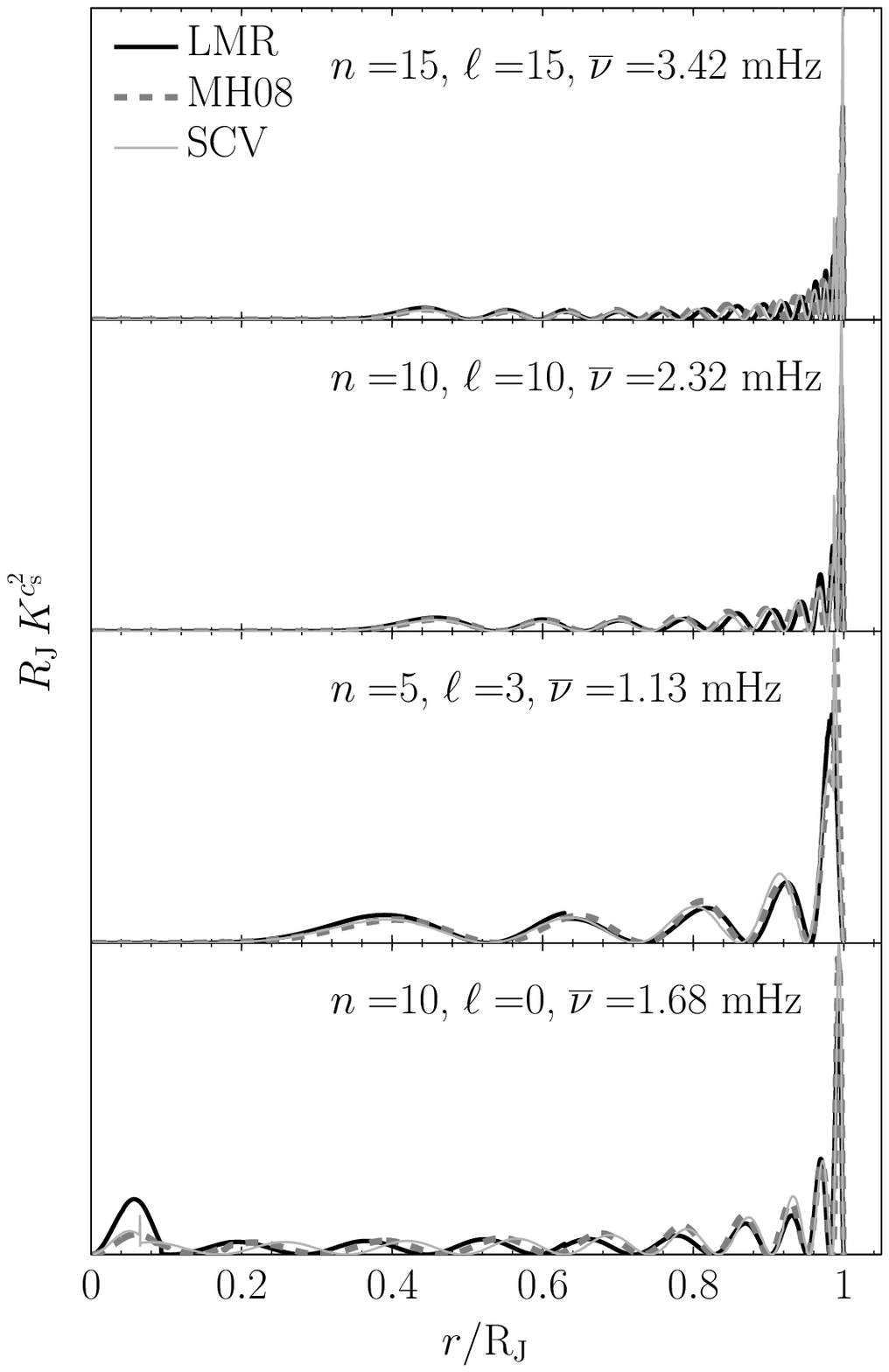}
    \includegraphics[width=.5\textwidth]{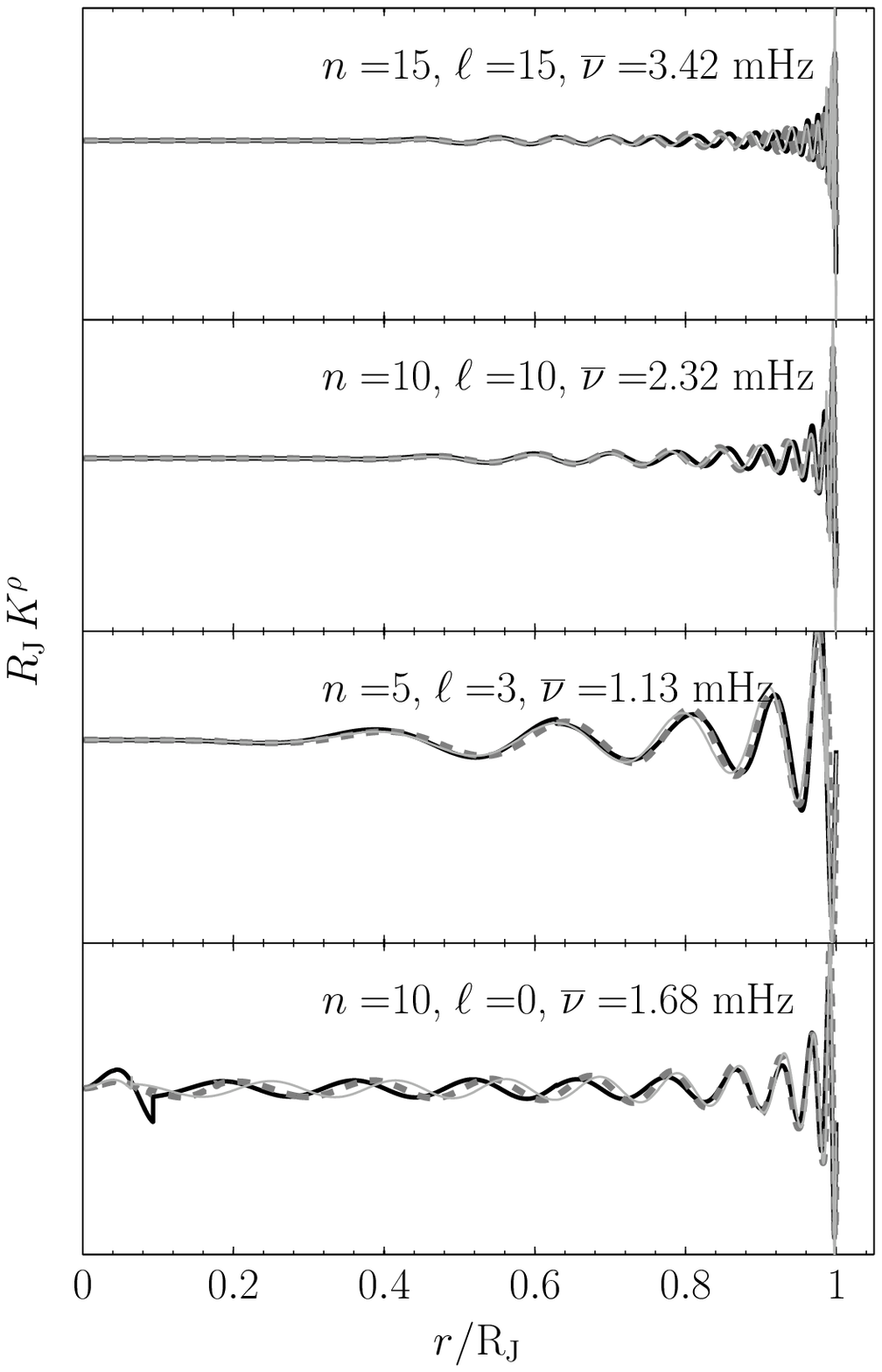}}
  \vspace{-.67\textwidth} 
  \centerline{\large\bf\hspace{0.06\textwidth}(a) \hspace{0.45\textwidth}(b)\hfill}
  \vspace{.6\textwidth}
  \caption{Sensitivity kernels for squared sound speed (a) and density (b) for the three Jovian models. Each panel represents a different angular degree, and various values for each of the radial orders were chosen for illustration purposes. Also labeled is the mean eigenfrequency $\overline{\nu}$ (averaged over each model) of the corresponding eigenfunctions used to construct the three kernels in each panel. Note the kernels are scaled with Jupiter's radius and are plotted in arbitrary units. The kernels in column (a) are positive and the kernels in (b) oscillate about zero.}
  \label{fig:kerns}
\end{figure}

Several example kernels for each of the three models are given in Fig.~\ref{fig:kerns}. Kernels for squared sound-speed are always positive, while those for density oscillate in sign. While overall these functions for each of the models are all quite similar, the LMR  kernels for low angular degree have strong core-boundary sensitivity. In general,  kernels for high-degree modes are less sensitive to inner regions, as expected from studying their lower turning points (Fig.~\ref{fig:rt}). We stress that the kernels are more useful than simply studying mode eigenfunctions, since they are computed for the specific observations in consideration: for this case these are frequency differences.

\section{Inversions of Jupiter frequencies}
\label{sec:inverse}

Assuming there are several available frequency measurements, the associated uncertainties, and a set of computed kernels, Eq.~\ref{csrho} has just two unknown quantities, $\delta \cs^2/\cs^2$ and $\delta\rho/\rho$. This is an inverse problem, that could in principle be solved by classical optimization methods, e.g., a least-squares algorithm. However, we emphasize that \textit{even in the absence of observations}, as is the case here, detailed analysis can be done upon ``inverting''  Eq.~\ref{csrho} to determine important predictive properties, such as the internal resolving power and expected errors of the final inferred quantities.

We  approach this problem by utilizing a Subtractive Optimally Localized Averages (SOLA) inversion procedure \citep[see e.g.][]{pijpers1992}, first developed for the case of the Sun. As shown below, it is a very useful tool even without available observations.  For completeness, we fully describe a one-dimensional version of this inversion procedure that is constructed to give the  most information at a given depth in Jupiter's interior. We follow with inversion examples using the Jovian models. 

\subsection{Formulation of the SOLA technique}

Qualitatively, the main goal of the SOLA inversion is to combine the kernels in such a way that a particular region of the Jovian interior is isolated. As is evident in Fig.~\ref{fig:kerns}, each kernel has a different depth sensitivity based on its frequency and angular degree. One can imagine that by adding and subtracting various weighted kernels, a desired \textit{averaging kernel} can be constructed that peaks at one particular depth, has tails that tend to zero, and is sufficiently narrow to provide a good spatial resolution. Ideally, a delta function would be preferred; however a finite set of oscillation modes and the presence of noise makes that impossible. The coefficients that are used for this optimized linear combination of individual kernels are then use to combine the relative frequency differences in the same  manner,  ultimately providing an estimate of the unknown interior quantity. The SOLA inversion finds such a set of optimal coefficients.

Assume $M$ number of perturbations $p$ (e.g., in this work $M=2$, for $p^1=\delta \cs^2/\cs^2$ and $p^2=\delta\rho/\rho$) and $N$ observations (i.e., $N$ frequency measurements of identified modes).  Equation~\ref{csrho} can be rewritten as 
\begin{equation}
  \frac{\delta\omega_i}{\omega_i} = \int_0^{\RJ}\sum_{\gamma=1}^M K^\gamma_i(r)\, p^\gamma(r)\,\id r + \epsilon_i,\;\;\; {\rm for }\;\; i=1 \ldots N,
  \label{csrho2}
\end{equation}
where the subscript $i$ corresponds to all $n\ell$ pairs of modes. We are interested in inferring a particular perturbation $p^\alpha$  at some radial depth $r_0$, and according to SOLA
\begin{equation}
  \overline{p}^\alpha(r_0) = \sum_{i=1}^N a_i^\alpha(r_0) \frac{\delta\omega_i}{\omega_i},
  \label{palpha}
\end{equation}
where  $\overline{p}^\alpha$ is an estimate of the true $p^\alpha$, and $\bvec{a}^\alpha$ is some set of coefficients yet to be determined. In other words, all available measurements are  averaged in some weighted fashion to infer the desired perturbation ($\alpha$) at a single depth ($r_0$). Plugging Eq.~\ref{csrho2} into Eq.~\ref{palpha} gives
\begin{eqnarray}\nonumber
  \overline{p}^\alpha(r_0) &=&  \sum_{i=1}^N a_i^\alpha(r_0) \int_0^{\RJ}\sum_{\gamma=1}^M K^\gamma_i(r)\, p^\gamma(r)\,\id r +  \sum_{i=1}^N a_i^\alpha(r_0)  \epsilon_i ,\\
&=&  \int_0^{\RJ}\!\! A^{\alpha,\alpha}(r,r_0)\,p^\alpha(r)\,\id r    +    \int_0^{\RJ}\!\!\!\sum_{\gamma=1,\gamma\neq\alpha}^M \!A^{\alpha,\gamma}(r,r_0) \, p^\gamma(r)\,\id r+  \sum_{i=1}^N a_i^\alpha(r_0)  \epsilon_i, \label{3terms}
\end{eqnarray}
where we have defined the  \textit{averaging kernel}
\begin{equation}
  A^{\alpha,\gamma}(r;r_0) = \sum_{i=1}^N a_i^\alpha(r_0)K_i^\gamma(r).
\end{equation}
Equation~\ref{3terms} is written in such a way as to reveal an important point: the estimate $\overline{p}^\alpha$ of the real $p^\alpha$ from the inversion is given by three terms. First, it is the convolution of  the true unknown quantity with the averaging kernel. If the averaging kernel is a well localized function centered around $r_0$, then this is a desirable result. The second contribution is from the contamination of  the other perturbations $\gamma\neq\alpha$ to our estimation, known as ``cross talk,'' and a large value is not desirable. Ideally, one would like $A^{\alpha,\gamma}(r)=0$. The third contribution is from the propagation of noise through the inversion.

To obtain the coefficient set $\bvec{a}^\alpha$, we minimize a cost function defined as
\begin{equation}
  X^\alpha(\mu) =  \int_0^{\RJ} \sum_{\gamma=1}^M\left[A^{\alpha,\gamma}(r;r_0) - T^{\alpha,\gamma}(r;r_0) \right]^2 + \mu\sum_{ij}^Na_i^\alpha(r_0)\Lambda_{ij}a_j^\alpha(r_0),
  \label{cost}
\end{equation}
where $T$ is a ``target'' function whose form is often taken to be 
\begin{equation}
  T^{\alpha,\gamma}(r;r_0) = \exp\left( \frac{-(r-r_0)^2}{2d^2}\right)\delta_{\alpha\gamma}.
\end{equation}
It is thus a Gaussian and peaked at a target depth $r=r_0$, where $d$ controls the width of the function. It is only nonzero for the perturbation being sought, as denoted by the Kronecker delta function $\delta$. The quantity $\Lambda_{ij}$  in Eq.~\ref{cost} denotes the noise covariance matrix of the measurements. 

Minimization of Eq.~\ref{cost} can therefore  be interpreted as a competition between  an averaging kernel that matches a prescribed target function and the contribution from noise. The quantity $\mu$ is a trade-off (free) parameter that allows one to tune between these two situations, depending on the type of solution desired. Ensuring a small ``misfit'' between the averaging kernel and the target can help localize around a particular depth, but will result in large error magnification, while relaxing that requirement gives more uncertainty and a coarser resolution, but  a smoother solution. Examples  will follow in the next section.

Finally, to allay the detrimental cross-talk contribution seen in Eq.~\ref{3terms}, we typically impose the constraint that 
\begin{equation}
  \int_0^{\RJ} A^{\alpha,\gamma}(r;r_0)\,\id r = \delta_{\alpha\gamma}.
  \label{constraint1}
\end{equation}
 Therefore, at least the total spatial integral of the undesired averaging kernels vanishes even though it typically does not vanish at each radial point $r$.   A second constraint is one that ensures the mass of Jupiter and the mass in the model are equal:
\begin{equation}
  4\pi\int_0^{\RJ}\frac{\delta\rho(r)}{\rho(r)}\rho(r)r^2\,\id r = 0.
\end{equation}
The minimization of Eq.~\ref{cost} with these two constraints leads to a system of $N+2$ linear equations  that are solved using well-known matrix methods.

\subsection{Inversion examples}

Despite the very promising observational analysis in \citet{gaulme2011} in recovering the large frequency separation, no measurements with unique Jovian mode identifications are yet available. We proceed with example inversions using   theoretical  sets of modes from each model for the ranges $\ell\leq 25$ and $0\leq\nu\leq 3.5$~mHz. The goal is to determine which observations would be most useful for inferring the sound speed and density profiles around several of the important interior regions of Jupiter, and  by quantifying the spatial resolution, noise estimation, and overall internal sensitivity expected from such modes. The analysis and conclusions assume that the given models are reasonably close to the true Jovian interior.

An estimation of measurement uncertainties is needed for the input to the inversion, and must therefore be approximated. We assume for simplicity that measurements are uncorrelated and have a relative uncertainty  $\epsilon_i=1\%$.  A diagonal noise covariance matrix of identical elements $\sigma^2$ will result, and the last term in Eq.~\ref{cost} reduces to $\mu\sum_i \sigma_i^2 a_i^2$. The choice of this conservative value is based on the fact that at least such precision is required to identify Jovian modes so  inversions can indeed be carried out. It corresponds approximately to $\Delta\nu/10$, and is  consistent with uncertainties provided in \citet{gaulme2011}.

\subsubsection{Targeting the outer-envelope transitions}

\begin{figure}
  \centerline{\includegraphics[width=\textwidth]{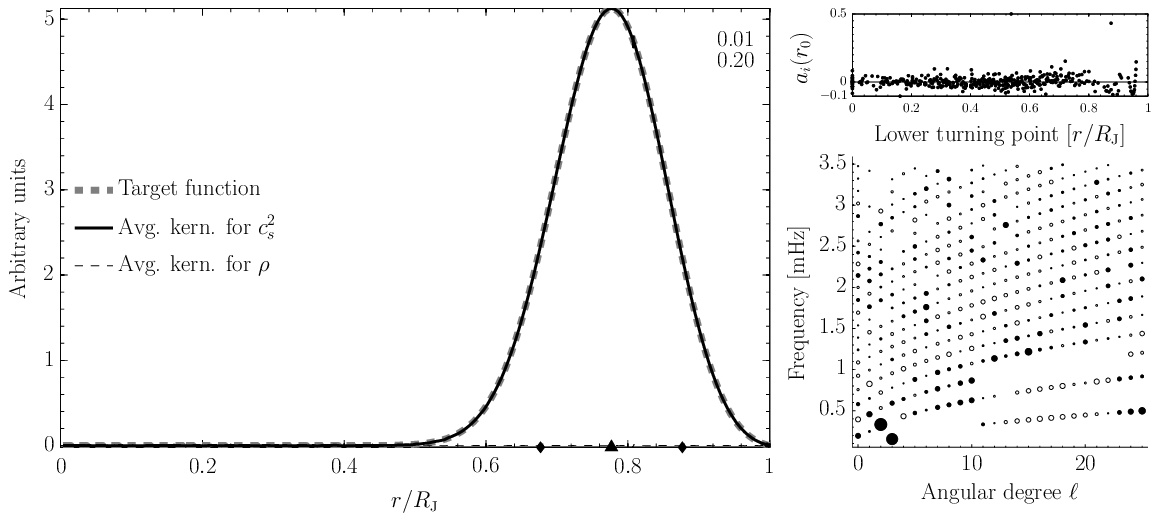}}
  \vspace{-.5\textwidth} 
  \centerline{\large\bf\hspace{0.07\textwidth}(a) \hspace{0.6\textwidth}(b)\hfill}
  \vspace{.46\textwidth}
  \centerline{\includegraphics[width=\textwidth]{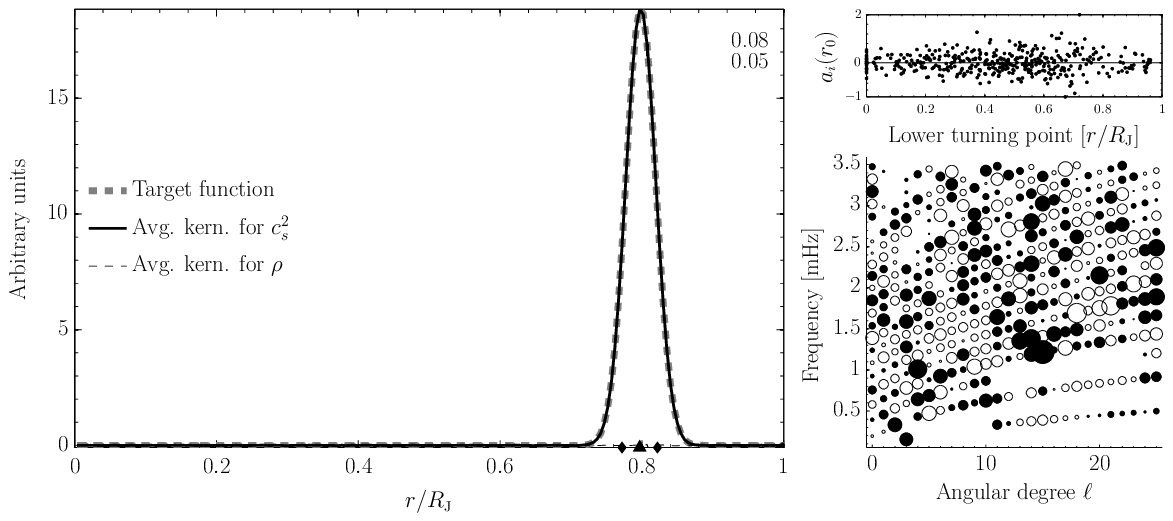}}
  \vspace{-.5\textwidth} 
  \centerline{\large\bf\hspace{0.07\textwidth}(c) \hspace{0.6\textwidth}(d)\hfill}
  \vspace{.4\textwidth}
  \caption{Example inversions for modes with $\ell\le 25$ using the SCV model to study  the sound-speed jump in the outer envelope transition of that model at $\sim 0.8\RJ$. Panels (a) and (b) show an inversion for a wide target function ($0.2\RJ$), and panels (c) and (d) for a narrower one ($0.05\RJ$).  On the $x$ axis, the triangle shows the target depth $r_0$, and the diamonds   $r_0\pm {\rm FWHM}/2$ of the prescribed target function. The upper right portion of panels (a) and (c)  provide the values of the inversion  relative error (in percent of squared sound-speed difference) and target FWHM (in fractional radius), respectively. Panels (b) and (d) show the properties of the coefficients $a_{n\ell}(r_0)$. The bottom plot in (b) and (d) denotes the relative strength of the coefficients by the size of the circles and the sign  (filled circles: positive, open circles: negative) in $\ell-\nu$ space. The top plot in these panels shows the amplitude of the coefficients as a function of their lower turning point.}
  \label{fig:scv_env}
\end{figure}

We first consider the outer layers of the models where the envelope transition occurs. For models SCV and LMR, the radial value of the transitions are at $\sim 0.8\RJ$ and $\sim 0.63\RJ$, respectively. Model MH08 has only a single envelope and will be used subsequently for studies of the core.

The jump in the SCV model sound-speed profile at $\sim 0.8\RJ$ is about $2.3\%$ (see Fig.~\ref{fig:cs}).  Figure~\ref{fig:scv_env} shows inversions for sound speed  targeting this transition region. Panel (a) shows the comparision of the two averaging kernels and the target function, centered at depth $r_0=0.8\RJ$, with a full-width at half-maximum (FWHM) of $0.2\RJ$. The averaging kernel for the sound speed matches  the target function very well, while the density perturbation averaging kernel is nearly zero throughout, thus eliminating any contributions of cross talk described earlier. Cancellation of sound-speed sensitivity is  achieved  in deep layers, while a small near-surface component remains.  As a reminder, in all inversions  the integral of the averaging kernel for sound speed is unity, while the integral of the density averaging kernel is zero, as constrained by Eq.~\ref{constraint1}.

The trade-off parameter in the inversion in Fig.~\ref{fig:scv_env}a (and all other inversions)   is chosen so that a good match with low noise  is obtained. This is done by inspecting an ``l-curve'' for the point of optimal error and misfit values \citep{jackiewicz2007a}. The combined noise after inversion is $0.01\%$. So even though we assume a $1\%$ noise level in each measurement, the optimized combination found in the inversion is much lower, as one would expect. 

The magnitude of the inversion coefficients $a_i(r_0)$ is shown in   Fig.~\ref{fig:scv_env}b as a function of  lower turning point $r_t$ to help visualize which modes contribute most to this particular inversion. We see that those whose $r_t$ lies near the target depth seem to have the largest amplitudes. We also plot the magnitude of the coefficients in an $\ell-\nu$ space representation. It is interesting to note that for some radial orders all modes along the ridge contribute with the same sign, while for other orders there are both positive and negative coefficients.

The second inversion shown in  Fig.~\ref{fig:scv_env}c and \ref{fig:scv_env}d is for a target function 4 times narrower, $0.05\RJ$, with everything else fixed. To achieve a reasonable match to the target function, however, we find the noise increases to about 8 times the value found for the inversion with the wider target. This is more evident in the plotted magnitudes of the inversion coefficients, and demonstrates  that it  requires more extreme weighting of the kernels  to find a  match to the narrow target.  Even such, the estimate of the sound speed at this target depth would have a standard deviation of only $\sigma\approx 0.04\%$ (note that the noise values provided in the figures are for the squared sound-speed differences $(c^2-c_0^2)/c^2$: the true fractional error is approximately found by halving the number),  making  a detection of a  2.3\% sound-speed jump as predicted by model SCV a realistic, although not a certain, goal.  However, to reduce the noise to lower value, a trade off with the the spatial resolution would be need to be made, and  more significant contributions from $A^\rho$ would factor into the inferences.


The conclusion of this example is that detecting the small 2.3\% jump in sound speed is reasonable to expect with this mode set based on noise levels, and the radial resolution appears to be sufficient  to pinpoint such a narrow feature. The example also demonstrates the rather subjective balance between spatial resolution and noise that must be carefully chosen.

\begin{figure}
 \centerline{\includegraphics[width=\textwidth]{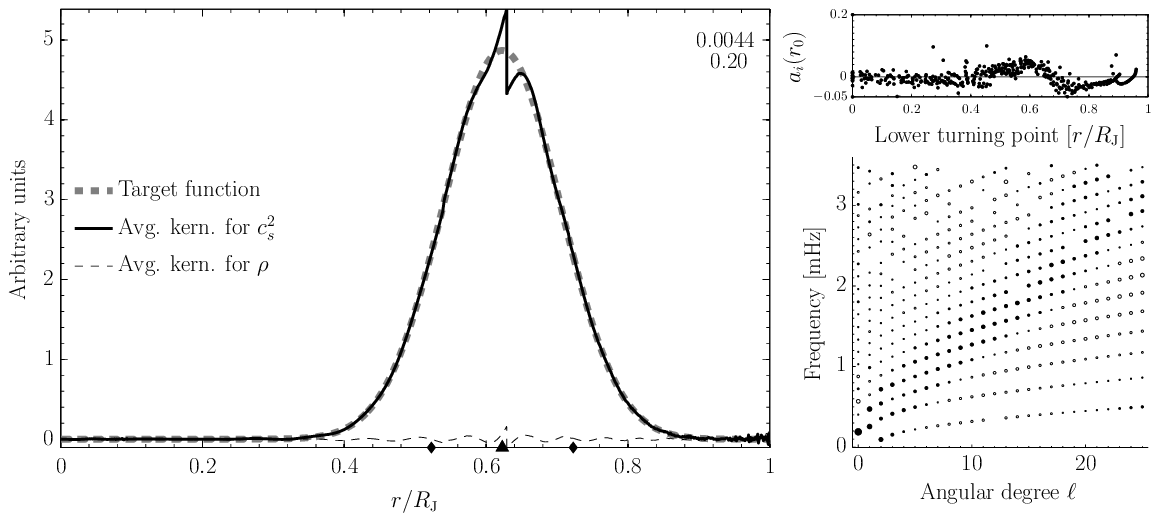}}
  \vspace{-.5\textwidth} 
  \centerline{\large\bf\hspace{0.07\textwidth}(a) \hspace{0.6\textwidth}(b)\hfill}
  \vspace{.46\textwidth}
  \centerline{\includegraphics[width=\textwidth]{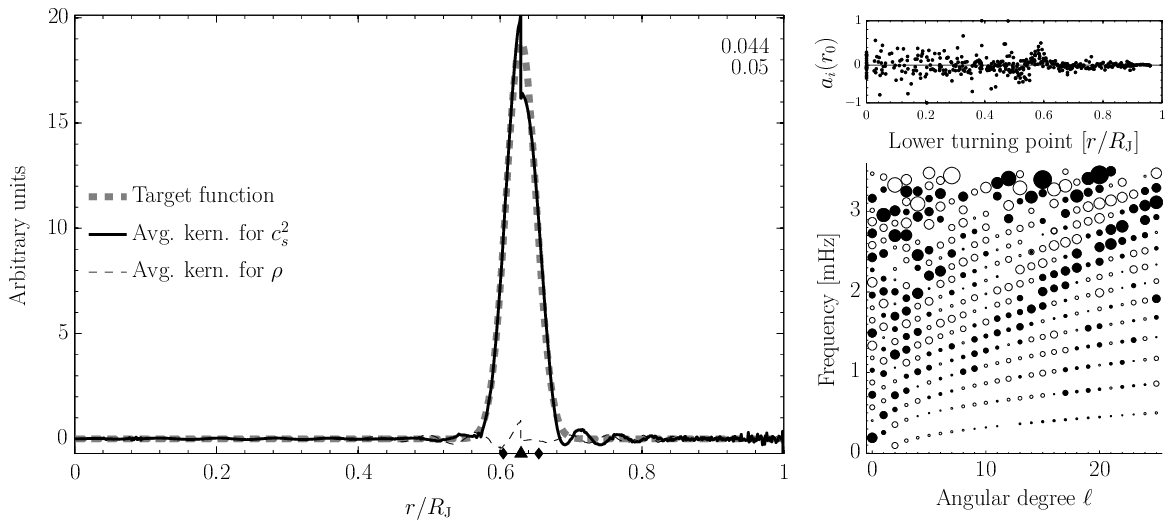}}
  \vspace{-.5\textwidth} 
  \centerline{\large\bf\hspace{0.07\textwidth}(c) \hspace{0.6\textwidth}(d)\hfill}
  \vspace{.4\textwidth}
  \caption{Example inversions using the LMR model for $\ell\le 25$  to invert for the sound-speed jump in the outer envelope transition of that model at $\sim 0.63~\RJ$. The figure follows the style of Fig.~\ref{fig:scv_env}.}
  \label{fig:lmr_env}
\end{figure}

Figure~\ref{fig:lmr_env} shows example inversions targeting the $\sim 5.5\%$ sound-speed jump in  model LMR at  $\sim 0.63\RJ$. The top panels present the results for a  wide target function of $0.2\RJ$. The sound-speed averaging kernel retains a sharp feature at this depth - there are not enough modes to ``average'' it away.  The noise is acceptable at 0.004\%, and the cross talk between sound speed and density again is minimal. Note the mostly positive sign of the coefficients for modes with lower turning point just beneath the target depth, and negative from those just above. {\em It is thus critical to have a full mode set to get this type of localization.}

For the inversion in Fig.~\ref{fig:lmr_env}c and \ref{fig:lmr_env}d with a narrower target and therefore better overall spatial resolution,  the noise is magnified by an order of magnitude from the previous case. For a sound-speed jump this large ($\sim~5.5\%$), however, the uncertainty ($0.2\%$) would not necessarily mask the detection. Given all of our assumptions, we conclude that this would be a detectable signal.

\begin{figure}
 \centerline{\includegraphics[width=\textwidth]{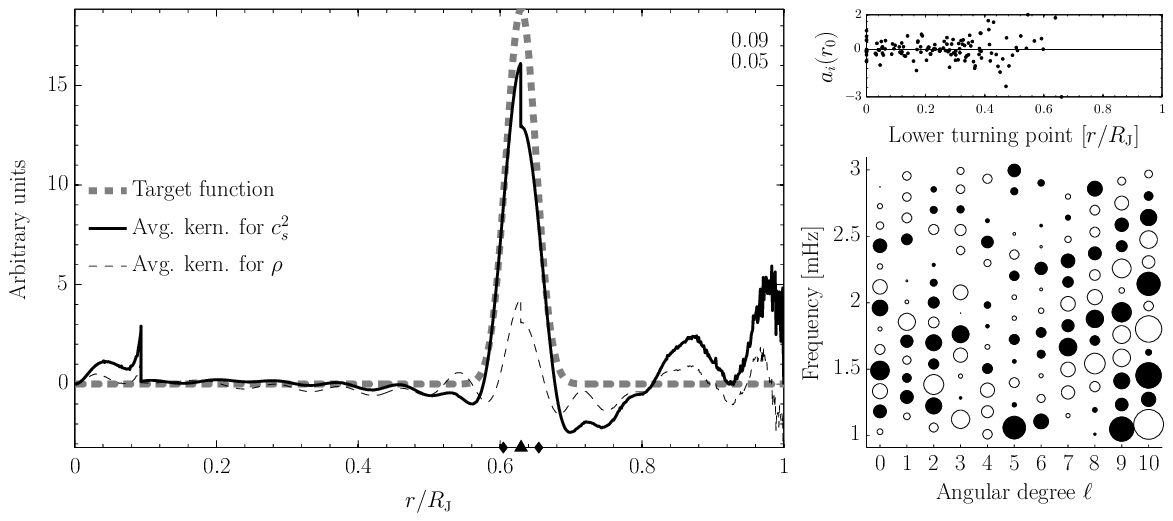}}
  \vspace{-.5\textwidth} 
  \centerline{\large\bf\hspace{0.07\textwidth}(a) \hspace{0.6\textwidth}(b)\hfill}
  \vspace{.4\textwidth}
  \caption{Example inversion using the LMR model with similar parameters as in Fig.~\ref{fig:lmr_env}c, but using a limited mode set for $\ell\le 10$  and $1\leq\nu\leq 3$ mHz. Note the much worse agreement.  This figure follows the style of Fig.~\ref{fig:scv_env}.}
  \label{fig:lmr_lim}
\end{figure}

To emphasize how important it is to have as large a set of observed modes as possible, we repeat the inversion discussed in  Fig.~\ref{fig:lmr_env}c with a more limited mode set. We constrain the input to modes of $\ell\leq 10$ and with frequencies between 1 and 3 mHz. The resulting inversion is shown in Fig.~\ref{fig:lmr_lim}. The degradation in the potential inference is striking, as the lack of a sufficient number of  high-degree modes greatly limits the inversion accuracy at this particular depth.

\begin{figure}
  \centerline{
    \includegraphics[width=.75\textwidth]{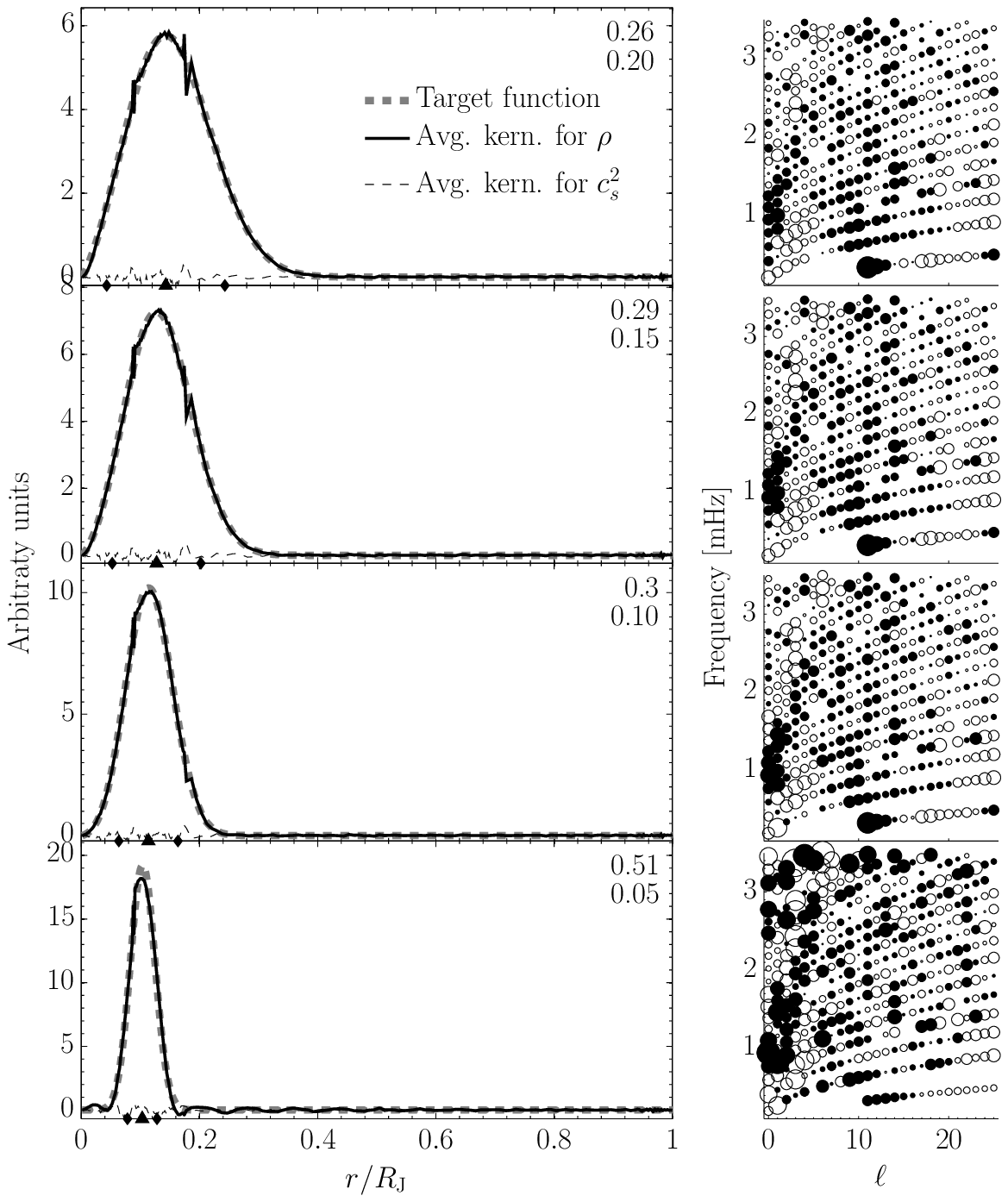}}
  \vspace{-.96\textwidth} 
  \centerline{\large\bf\hspace{0.2\textwidth}(a)\hspace{0.45\textwidth}(b)\hfill}
  \vspace{.86\textwidth}
  \caption{Example inversions for density differences using model MH08 to target the core     boundary layers at about $r_0=0.1\RJ$. The panels in column (a) show comparisons of the averaging kernels with the target function for varying FWHM, decreasing from top to bottom. The panels in column (b) show the corresponding contributions of the  inversion coefficients in $\ell-\nu$ space.}
  \label{fig:inv_hub1}
\end{figure}

\begin{figure}
  \centerline{
    \includegraphics[width=.75\textwidth]{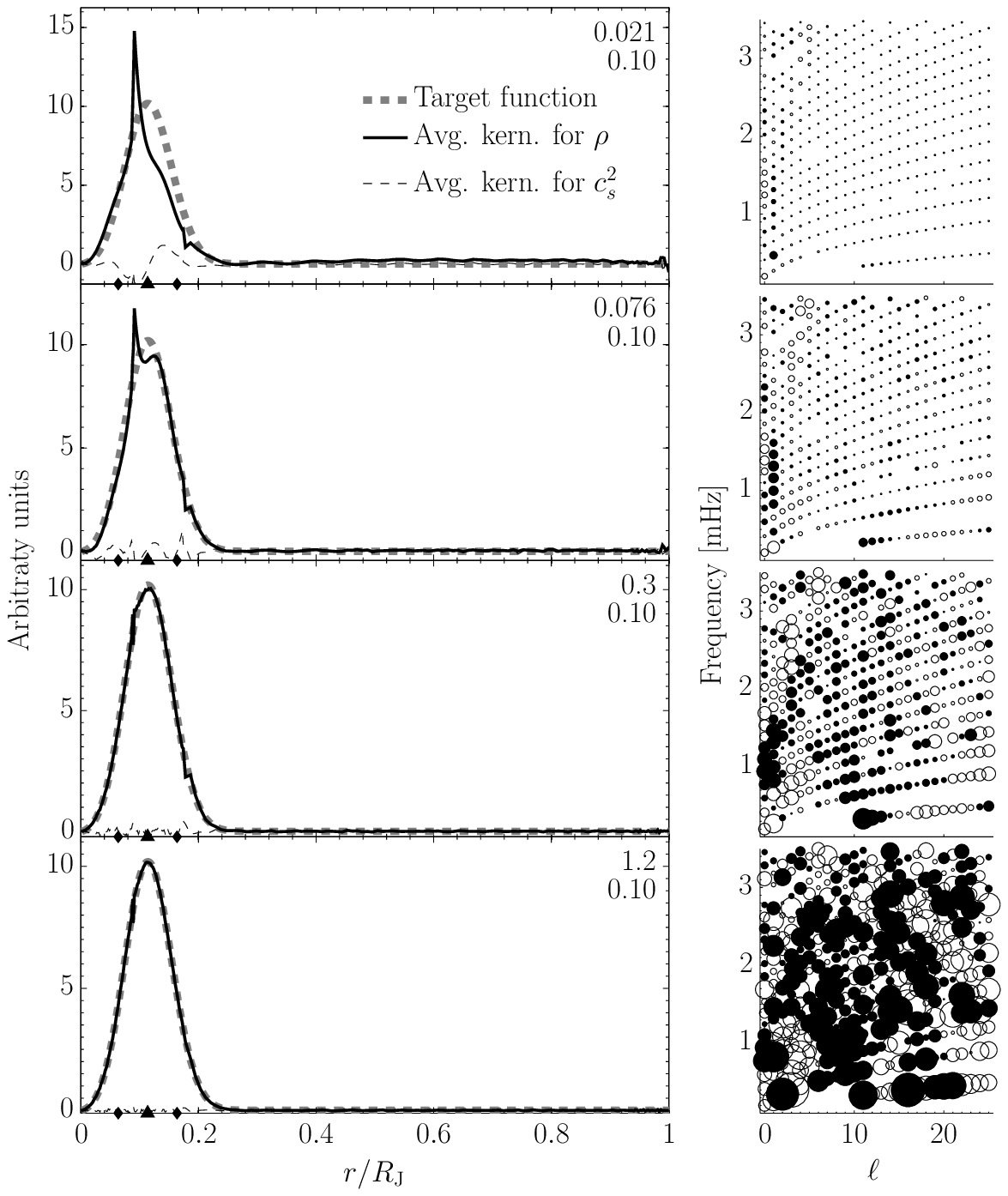}}
  \vspace{-.96\textwidth} 
  \centerline{\large\bf\hspace{0.2\textwidth}(a)\hspace{0.45\textwidth}(b)\hfill}
  \vspace{.86\textwidth}
  \caption{Example inversions for density differences using model MH08 to target the core   boundary layers as in Fig.~\ref{fig:inv_hub1}, but varying the trade-off parameter. The panels in column  (a) shows inversions with fixed FHWM$=0.1$, but for $\mu$ decreasing from top to bottom by two orders of magnitude in each panel. Column (b) plots the coefficients for these inversions  in $\ell-\nu$ space. }
\label{fig:inv_hub2}
\end{figure}

\subsubsection{Targeting the core-boundary transition}

We shift the focus to the core-envelope boundary and study  the massive-core ($16\ME$) MH08 model. From Fig.~\ref{fig:cs} there is an overall $\sim 80\%$ drop in the density from the core region to the envelope. While this should produce a strong seismic signal, there are indeed fewer modes that probe to such depths.

Figure~\ref{fig:inv_hub1} shows inversions for density in model MH08 around $r_0=0.1\RJ$. We compare a series of inversions in which the FWHM of the target function decreases from top to bottom panels.  We verify from the plots of the coefficients  that the low-degree modes contribute most to the averaging at this depth. It is precisely because of the lack of a large number of low-degree modes that causes  some  difficulty in resolving the deep interior regions. There are also small but possibly significant contributions from cross talk with the sound speed, shown by the oscillating features in the deep interior.

We next perform inversions after choosing a fixed target function with ${\rm FWHM}=0.1\RJ$, and sweep through several values of the trade-off parameter $\mu$ to study its effect on the inversion.   Figure~\ref{fig:inv_hub2} shows the corresponding inversions in this case.  The value of $\mu$ decreases by two orders of magnitude in each panel from top to bottom -  large $\mu$ means more emphasis on obtaining low noise, while  smaller values typically gives  better fits of the averaging kernel to the target, as Eq.~\ref{cost} describes. This trend is evident in the averaging kernels for density as the target/kernel match improves in each successive panel, while the noise increases as well.

The maximum noise reaches about 1.2\%, which is larger than the value of each individual measurement (1\%).  However, the spatial resolution is quite high and the cross talk is minimal for that particular inversion (even though it is clear from the amplitudes of the weights that most of the modes are weighted very strongly compared to the other examples). For a deep density jump of 80\%, we would therefore expect to detect its location with this small set of modes,  even though these  uncertainties might not allow a fully accurate determination of its overall amplitude.

We finally note that in these deep-layer inversions the high-degree modes are useful even though they do not probe deeply and only contribute rather small-amplitude weighting as Fig.~\ref{fig:inv_hub2} shows. We have tested similar inversions using only $\ell\leq 10$, and find that the fit worsens due to the fact that near-surface cancellation does not occur as efficiently than with these modes and non-local effects away from $r_0$ are significant. Overall, it is likely advantageous to use as many modes as possible in each inversion, regardless of target depth.


%

\section{Conclusions}
\label{sec:conc}

Recent seismic observations and  analyses have indicated that the detection of pulsation modes of Jupiter is now possible, and the large frequency separation between low-degree modes of consecutive radial order has been measured. In this work we study three recent state-of-the-art models of the Jovian interior that span a broad range of parameter space. We assume a realistic acoustic oscillation mode set that, in principle, could be obtained from current ground or space-based instrumentation. This set contains angular degrees $\ell\leq 25$ and frequencies below the approximate acoustic-cutoff frequency of Jupiter of $\nu\leq 3.5$~mHz. We compute these modes using well-known  methods, and find that the theoretical frequencies are consistent with past studies. 

 For the first time, we show theoretical sensitivity kernels computed from Jupiter models and inversions for select regions of the interior. We use forward and inverse techniques developed for helioseismology. Using the example modes, we invert for  sound speed and density perturbations.

The three models exhibit  discontinuities at interior boundary layers, and  we conclude that we could reasonably detect sound-speed jumps of a few percent in the inner envelope if the modes studied here were observed and available.  For deep transitions in the core boundary where fewer modes probe, it is still very likely that the locations of the large discontinuities could be accurately found, but the magnitudes of the jumps will be somewhat more difficult to ascertain because of noise. All modes have  different sensitivities at different depths, and these results  depend on a number of factors, including the parameters used in any inversion. While a set of simplifying assumptions was necessary to perform this analysis, these examples do provide a powerful demonstration of what to expect once confirmed measurements are available.

New dedicated observations of Jovian oscillations are strongly encouraged based on these results, which show that most of the interesting features of Jupiter's interior are captured by inverse techniques. We strongly encourage  observers to design a strategy to detect, at bare minimum, the mode set studied here. For a detailed study of the outer layers of the planet near the surface, even higher-degree modes that are trapped in that region would be necessary to measure and identify.


We thank Patrick Gaulme for very useful comments about the paper, and acknowledge the suggestions by two anonymous referees that made the text significantly clearer. This work was supported by the NASA Outer Planets Research Program.

\bibliographystyle{plainnat}
\bibliography{myrefs}

\end{document}